\documentclass[twocolumn,superscriptaddress,showpacs,aps,floatfix]{revtex4-1}
\usepackage{amsmath,amsfonts,amssymb,epsfig,dcolumn,bm,dsfont,graphics,latexsym,color,graphicx,multirow,ulem}
\usepackage[bookmarks,colorlinks,linkcolor=blue,citecolor=blue,urlcolor=blue]{hyperref}
\newcommand{\ket}[1]{| {#1} \rangle} 
\newcommand{\bra}[1]{\langle {#1} |} 

\newcommand{\da}[1]{\downarrow}
\newcommand{\ua}[1]{\uparrow}

\begin{document}
\preprint{AIP/123-QED}
\title{Proximity-induced unconventional superconductivity and chiral topological 
phases in twisted graphene/NbSe$_2$ van der Waals heterostructure}

\newcommand{\SAVBA}{\affiliation{Institute of Informatics, Slovak Academy of Sciences, 84507 Bratislava, Slovakia}}
\newcommand{\FIT}{\affiliation{Faculty of Informatics and Information Technologies, Slovak University of Technology, 842 16 Bratislava 4, Slovakia}}
\newcommand{\FFBG}{\affiliation{Faculty of Physics, University of Belgrade, 11001 Belgrade, Serbia}}

\author{Adam Hlo{\v z}n{\' y}} \SAVBA \FIT
\author{Marko Milivojevi{\' c}} \SAVBA \FFBG

\begin{abstract}
We study proximity-induced unconventional superconductivity in a twisted graphene/NbSe$_2$ van der Waals heterostructure using the Bogoliubov-de Gennes formalism. The normal-state parameters of 
proximitized graphene are extracted from ab initio 
calculations at a twist angle of $23.4^\circ$, which reduces the common symmetry of the heterostructure to $\mathbf{C}_3$. We construct symmetry-allowed superconducting gap functions of the graphene layer according to the irreducible 
representations of the $\mathbf{C}_3$ group, containing singlet and triplet pairing channels and their mixtures. Computing the topological invariants as a function of the mixing parameters, we find a rich phase diagram of chiral topological superconducting phases, characterized by nonzero  Chern numbers $C\in\{-4,-2,2,4\}$. While the nature of the superconducting order parameter of NbSe$_2$ remains debated, the formation of the van der Waals 
heterostructure and the related symmetry reduction can alter the relative stability of competing pairing channels, potentially stabilizing a chiral component that is proximity-induced into graphene and triggers the topological phases identified here, making the twisted graphene/NbSe$_2$ heterostructure a promising platform for chiral topological superconductivity detectable via quasiparticle interference imaging and transport measurements.
\end{abstract}
\maketitle
\section{Introduction}
Spintronics~\cite{ZFS04,FME+07} is a subfield of electronics that focuses on the spin instead of the charge as the active degree of freedom for storing, processing, and transmitting information.
Spintronics effects, such as charge-spin conversion~\cite{Offidani2017,Ghiasi2019}, 
the spin Hall effect~\cite{Sinova2015}, and spin transport~\cite{TJP+07}, open routes toward devices with lower power dissipation and new functionalities beyond charge-based electronics.

Graphene~\cite{NGM+04} represents a material interesting for spintronics application due to its long spin diffusion lengths~\cite{HGN+06,TJP+07} and high mobilities~\cite{NGM+04}. These properties enable spin injection~\cite{OSN+07}, transport~\cite{NGM+04}, and detection~\cite{GOW17,GOW18}, as demonstrated in devices like spin valves~\cite{HGN+06} and field-effect transistors~\cite{Popinciuc2009,Ringer2018}. However, the weak intrinsic spin-orbit coupling (SOC) of graphene limits its applicability for active spin manipulation. By placing graphene in a van der Waals heterostructure with a material possessing strong SOC, this limitation can be overcome. Due to the spin-orbit proximity effect, graphene acquires sizable SOC while keeping its transport properties~\cite{Avsar2014,Gmitra2015}. Thus, the proximity-induced SOC  enables manipulation of spin 
currents~\cite{YTL+16} and charge-spin conversion~\cite{CSC+22,Lee2022b,OSH+23,Chi2024,CDY+25,MMJ+26} without the need for external magnetic fields, offering efficient 
spin control at the nanoscale.

Graphene can also become superconducting through the superconducting proximity effect when placed in contact 
with a superconductor, representing a platform for superconducting spintronics~\cite{ALR+24,MHL+20}. In recent years, it has become clear that the spin-orbit proximity effect is important in stabilizing and enhancing superconductivity in graphene-based 
heterostructures~\cite{ZPT+23}. For example, it was recently shown~\cite{ZSM+25} that Bernal bilayer graphene proximitized by tungsten diselenide exhibits twist-programmable superconductivity. More concretely, it was shown the relative twist angle between the two layers controls the strength of the proximity-induced SOC, which in turn enhances the critical temperature and the upper critical field, demonstrating that the twist angle represents an experimental knob for engineering tunable graphene-based superconductors.

In this context, the twisted graphene/NbSe$_2$ heterostructure is a 
particularly promising platform. NbSe$_2$ is a layered superconductor with a bulk critical  temperature of $T_c \approx 7.1$~K~\cite{Sanchez1995}, that is reduced in the monolayer limit~\cite{Xi2016,
Khestanova2018}, and simultaneously possesses strong SOC. When graphene is placed on 
NbSe$_2$, two distinct proximity effects operate simultaneously: 
the spin-orbit proximity effect, which induces sizable intrinsic and Rashba 
SOC in the graphene layer~\cite{HFK+20,NGF24}, 
and the superconducting proximity effect, which induces Cooper pairing 
directly into  graphene~\cite{NMA+25,TKO+25}. The coexistence 
of these two proximity effects makes the graphene/NbSe$_2$ heterostructure 
an ideal platform for studying the interplay of proximity-induced SOC and superconductivity in two dimensions. 


In this work, we study proximity-induced superconductivity in a twisted graphene/NbSe$_2$ heterostructure using the Bogoliubov-de-Gennes (BdG) formalism. The normal-state electronic structure of proximitized graphene is described by an effective tight-binding Hamiltonian incorporating orbital hopping, intrinsic SOC, and Rashba SOC, with parameters  determined from density functional theory (DFT) calculations of the heterostructure at a twist angle of $23.4^\circ$.
The superconducting gap function entering the BdG Hamiltonian is constructed using group theory, with the common symmetry between graphene and NbSe$_2$ being reduced to ${\bf C}_3$ due to the finite twist angle. We classify symmetry-allowed superconducting order parameters according to the irreducible representations (IRs) $A_0$, $A_1$, and $A_{-1}$ of the ${\bf C}_3$ group. Within this approach, the singlet and triplet pairing channels are treated on equal footing, since symmetry-based construction naturally allows for singlet-triplet mixing. 
Using the derived gap functions together with the DFT-based parameters of the proximitized graphene, we study the topological properties of the superconducting graphene as a function of the mixing angles parametrizing the singlet-triplet 
and inter-triplet(singlet) mixing of the order parameter. Our phase diagram reveals different topological superconducting phases driven by unconventional pairing that breaks the time-reversal symmetry (TRS), suggesting that the graphene/NbSe$_2$ heterostructure is a promising platform for proximity-induced chiral and topological superconductivity.

The paper is organized as follows. In Section~\ref{SectionII} we introduce the BdG formalism and the effective tight-binding model of proximitized graphene with ${\bf C}_3$ symmetry, alongside with the DFT-derived parameters. In Section~\ref{GapContruction}, group-theoretical construction of symmetry-allowed superconducting gap functions in the on-site and nearest-neighbor approximations is provided. In Section~\ref{topology}, the phase diagrams of superconducting phases are analyzed. The conclusions are given in Section~\ref{CONC}.

\begin{figure}[t]
\centering
\includegraphics[width=0.495
\textwidth]{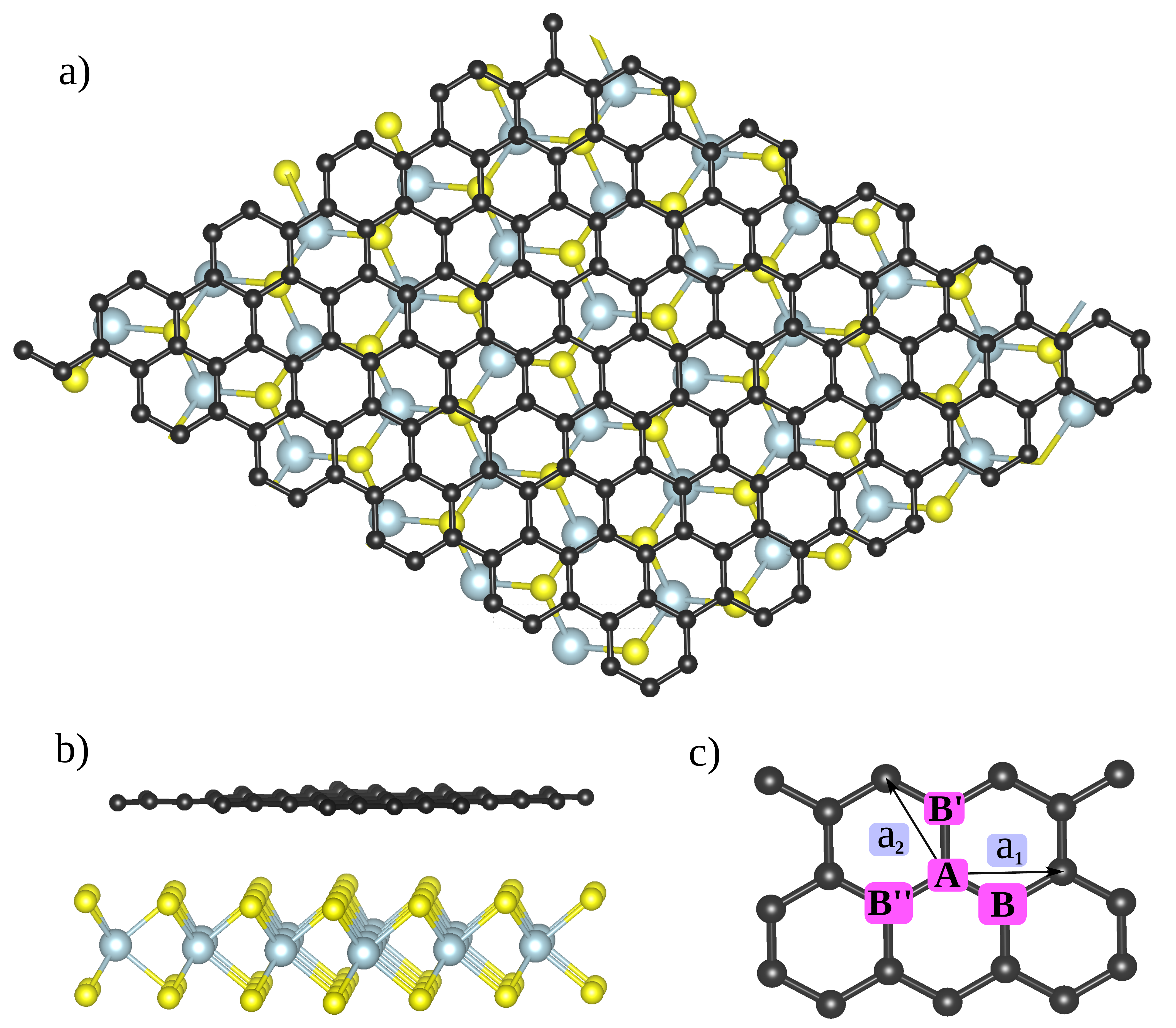}
\caption{Top (a) and side (b) view of the graphene/NbSe$_2$ heterostructure with the twist angle $\theta=23.4^{\rm o}$  between the graphene and NbSe$_2$ monolayer. (c) Position of graphene sublattice A and B carbon atoms and the graphene lattice vectors ${\bf a}_1=a(1,0,0)$ and
    ${\bf a}_{2}=a(-\frac{1}2,\frac{\sqrt{3}}2,0)$ used to derive the effective model of graphene~\eqref{Hgr} in its normal state. In addition to this, nearest neighbor atoms of carbon A atom are depicted, used to construct the superconducting gap functions in Section~\ref{nnAPPROX}. }\label{FIG1}
\end{figure} 

%
\section{Effective BdG model for proximitized graphene}\label{SectionII}

We are interested in studying the effects of electron superconducting pairing in proximitized graphene with ${\bf C}_{3}$ symmetry at a low-temperature regime. Superconductivity is modeled using the BdG formalism with Hamiltonian 
\begin{equation}
    \mathcal{H}_\mathrm{BdG} = \sum_\mathbf{k} \Psi^\dagger_\mathbf{k} \mathcal{H}_{\mathbf{k}}^\mathrm{BdG} \Psi_\mathbf{k},
\end{equation}
writen in terms of the Nambu spinor, $\Psi^\dagger_{\mathbf{k}} = [ a^\dagger_{\mathbf{k}\uparrow}, a^\dagger_{\mathbf{k}\downarrow},b^\dagger_{\mathbf{k}\uparrow}, b^\dagger_{\mathbf{k}\downarrow},
a_{\mathbf{-k}\uparrow}, a_{\mathbf{-k}\downarrow},b_{\mathbf{-k}\uparrow}, b_{\mathbf{-k}\downarrow}]$ with fermionic creation and annihilation operators $a^{\dag}/a$ and $b^{\dag}/b$ on sublattice A and B, respectively (see FIG.~\ref{FIG1}c). The $8\times 8$ BdG Hamiltonian $\mathcal{H}_{\mathbf{k}}^{\mathrm{BdG}}$ in the reciprocal space can be writen following the procedure discussed in~\cite{AZ97} and taking into the account that electrons and holes have opposite ${\bf k}$ momentum in the Nambu basis
\begin{equation}\label{EQ:BDG}
    \mathcal{H}_{\mathbf{k}}^{\mathrm{BdG}} =
    \begin{pmatrix}
        \mathcal{H}_\mathrm{e}(\mathbf{k}) & \Delta (\mathbf{k})\\
        \Delta^{\dagger}(\mathbf{k})& -\mathcal{H}_\mathrm{e}^{\rm T}(-\mathbf{k})
    \end{pmatrix}.
\end{equation}
In the previous equation, the electron Hamiltonian in ${\bf  k}$ space, $\mathcal{H}_\mathrm{e}(\mathbf{k})$, represents the effective tight-binding Hamiltonian of proximitized graphene with ${\bf C}_{3}$ symmetry~\cite{David2019}
\begin{equation}\label{Hgr}
    \mathcal{H}_\mathrm{e}(\mathbf{k}) =  \mathcal{H}_\mathrm{orb}(\mathbf{k})+ \mathcal{H}_\mathrm{I}(\mathbf{k}) + \mathcal{H}_\mathrm{R}(\mathbf{k}),
\end{equation}
where the orbital part $\mathcal{H}_\mathrm{orb}(\mathbf{k})$ describes the dispersion of the carbons $p_z$-band, while $\mathcal{H}_\mathrm{I}(\mathbf{k})$ and $\mathcal{H}_\mathrm{R}(\mathbf{k})$ represent the intrinsic and Rashba SOC contributions.
After constructing $\mathcal{H}_{\rm e}({\bf k})$, the hole Hamiltonian $\mathcal{H}_{\rm h}({\bf k})$ is straighforwardly obtained, since it is defined as $\mathcal{H}_{\rm h}({\bf k})=-\mathcal{H}_{\rm e}(-{\bf k})^{\rm T}$ in the basis writen above. The final part of the BdG Hamiltonian represent the superconducting gap function $\Delta({\bf k})$, whose structural form is not known, but the possible forms are restricted by the symmetry of the system under study. The derivation of possible superconducting pairing functions will be done in the next section.

Before that, we recapitulate the effective tight-binding model of graphene with ${\bf C}_{3}$ symmetry, which can be used to derive its Bloch Hamiltonian form $\mathcal{H}_e({\bf k})$. We start from the Hamiltonian
$\mathcal{H}_{\rm orb}$.
After defining the effective $p_z$-orbital with spin $\sigma=\uparrow,\downarrow$ on sublattice $X=A,B$ on lattice site $m$ as $\ket{X_m\sigma}$, $\mathcal{H}_{\rm orb}$ can be written as
\begin{eqnarray}
\mathcal{H}_\mathrm{orb}&=&\sum_{m,\sigma}\mu_A\ket{A_m\sigma}\bra{A_m \sigma}+\sum_{m,\sigma}\mu_B\ket{B_m\sigma}\bra{B_m \sigma}\nonumber\\
&&-t\sum_{<{m,n}>_{\rm nn}}\sum_{\sigma}\ket{X_m\sigma}\bra{X_n \sigma},
\end{eqnarray}
described in terms of the sublattice-dependent on-site  potential,
equal to $\mu_{A/B}=\mu\pm \delta$ on sublattice A/B of graphene, where $\mu$ is the chemical potential, and $\delta$ is the stagerred potential.
Additionally, orbital Hamiltonian contains the nearest-neighbor interaction, parametrized by the coupling strength $-t$.

The second term in Eq.~\eqref{Hgr} in the real space
\begin{equation}
    \mathcal{H}_{\rm I}=\sum_{\gamma={\rm A,B}}\sum_{<m,n>_{\rm nnn},\sigma}\frac{{\rm i}\lambda_{\rm I}^{\gamma}}{3\sqrt{3}}\nu_{m,n}[s_z]_{\sigma\sigma}\ket{X_m\sigma}\bra{X_n \sigma},
\end{equation}
describes the intrinsic SOC, where the sum goes over the next-nearest-neighbors interaction described in terms of the sublattice-dependent $\lambda_{\rm I}^{\rm{A}/\rm{B}}$ parameters and the sign factor $\nu_{m,n}$ that has the value 1 ($-1$) when the next-nearest-neighbor hopping from site $m$ to site $n$ via the common nearest-neighbor encloses a clockwise (counterclockwise) path~\cite{Kochan2017}. 

\begin{table}[t]
\caption{\label{tab:CDW_ex}
Effective parameters of graphene, obtained by fitting the effective graphene model~\eqref{Hgr}  
to the DFT bands and spin expectation values of graphene (within the twisted graphene/NbSe$_2$ heterostructure) in the vicinity of the Dirac $K$ point. The effective parameters of the graphene Hamiltonian, such as chemical potential $\mu$, hopping parameter $t$, staggered potential $\delta$, intrinsic spin-orbit coupling parameters $\lambda_{\mathrm {I}}^A$,  $\lambda_{\mathrm {I}}^B$, and the Rashba parameter $\lambda_{\rm R}$
are given in meV units, whereas the
Rashba angle $\phi_{\rm R}$ is given in degrees.}
\centering
\begin{tabular}{c|c|c|c|c|c|c} 
$\mu$ &$t$  &$\delta$ & $\lambda_{\mathrm {I}}^A$  & $\lambda_{\mathrm {I}}^B$  & $\lambda_{\mathrm R}$  & $\phi_{\mathrm R}$ (deg)\\\hline
525.045&2550.67&-0.595&0.067&-0.29&-1.496&-22.975 \\
\end{tabular}
\end{table}

The last term in Eq.\eqref{Hgr} describes the nearest-neighbor Rashba SOC interaction  
\begin{equation}
\mathcal{H}_{\rm{R}}=\frac{2{\rm i}
\lambda_{\rm R}}{3}\sum_{<m,n>_{\rm nn}}^{\sigma\neq\sigma'}
U^{\dagger}_{\phi_R}[{\bf s}\times {\bf d}_{m,n}]_{\sigma\sigma'}^zU_{\phi_R} \ket{X_m\sigma}\bra{X_n \sigma'},
\end{equation}
in which ${\bf s}$ is the vector of Pauli matrices, $\lambda_{\rm R}$ represents the Rashba SOC strength, while ${\bf d}_{m,n}$ is the unit vector in the horizontal plane pointing from lattice site $n$ to the nearest-neighbor site $m$. Finally, the unitary operator $U_{\phi_{\rm R}}={\rm e}^{{\rm i}\sigma_z \phi_{\rm R}/2}$ appears due to the ${\bf C}_3$ symmetry of the heterostructure (nonzero twist angle)~\cite{David2019}. 

The unknown parameters in $\mathcal{H}_{e}({\bf  k})$ are the chemical potential $\mu$, hopping parameter $t$, staggered potential $\delta$, intrinsic spin-orbit coupling parameters $\lambda_{\mathrm {I}}^A$ and $\lambda_{\mathrm {I}}^B$, as well as the Rashba parameter $\lambda_{\rm R}$ and the Rashba angle $\phi_{\rm R}$. 
To study the superconducting properties using realistic parameters of graphene in the normal phase, we performed DFT calculation of the twisted graphene/NbSe$_2$ heterostructure, having the twist angle of 23.4$^{\rm o}$ between graphene and NbSe$_2$ (see FIG.~\ref{FIG1}a and FIG.~\ref{FIG1}b as an illustration, whereas more details can be found in Appendix~\ref{AppA}).
By fitting $\mathcal{H}_{e}({\bf  k})$ to the DFT-obtained energy dispersion and spin expectation values in the vicinity of the Dirac K point of graphene, we obtained the effective parameters of graphene, see Table~\ref{tab:CDW_ex}. Our results are in good agreement with~\cite{NGF24}, with some small discrepancies that are consequence of different strategy used in DFT calculation: whereas in our work the graphene/NbSe$_2$ heterostructure is fully relaxed, in~\cite{NGF24} the relaxation calculation was not performed.
\section{Constructing the superconducting gap function}\label{GapContruction}
The final part of the BdG Hamiltonian represents the superconducting gap function $\Delta({\bf k})$. For monolayer graphene, one can write the gap function $\Delta({\bf k})$ as
\begin{equation}
\Delta({\bf k})=    \Big(\begin{array}{c|c}
     \Delta^{\rm AA}_{\bf k}    & \Delta^{\rm AB}_{\bf k}  \\\hline
     \Delta^{\rm BA}_{\bf k}   &  \Delta^{\rm BB}_{\bf k}
    \end{array}\Big),
\end{equation}
where all the gap components  $\Delta^{\rm AA/BB/AB/BA}_{\bf k}$
are the $2\times2$ blocks that can be spanned using the $d$-matrices $d_0={\rm i}\sigma_y$, $d_{x/y/z}={\rm i}\sigma_{x/y/z}\sigma_y$, where $d_0$ describes the singlet pairing $(\ket{\uparrow\downarrow}-\ket{\downarrow\uparrow})$ , while $d_{x/y/z}$ correspond to the different types of triplet spin pairing functions:
$d_x\sim(-\ket{\uparrow\uparrow}+\ket{\downarrow\downarrow})$, 
$d_y\sim(\ket{\uparrow\uparrow}+\ket{\downarrow\downarrow})$, and
$d_z\sim(\ket{\uparrow\downarrow}+\ket{\downarrow\uparrow})$.
The presence of the ${\bf C}_{3}$ symmetry fully decouples the $d_z$ matrix from the two-dimensional $(d_x,d_y)$ subset, while $d_0$ matrix is naturally singled out as the only singlet component. Furthermore, anti-symmetry condition on the gap parameter due to the Pauli principle implies that $\Delta^{A\sigma,A\sigma'}_{\bf k}= -\Delta^{A\sigma',A\sigma}_{-{\bf k}}$, $\Delta^{B\sigma,B\sigma'}_{\bf k}= -\Delta^{B\sigma',B\sigma}_{-{\bf k}}$
$\Delta^{A\sigma,B\sigma'}_{\bf k}= -\Delta^{B\sigma',A\sigma}_{-{\bf k}}$. Thus, it is enough to determine $\Delta^{\rm AB}_{\bf k}$; the Pauli principle automatically gives the other $\Delta^{\rm BA}_{\bf k}$ block. On the other hand, the antisymmetry condition implies that the channels $\Delta^{\rm AA}_{\bf k}$ and $\Delta^{\rm BB}_{\bf k}$ themselves must satisfy the anti-symmetry condition.

Since the ${\bf C}_3$ symmetry ot the heterostructure keeps the distance between the
atoms fixed, one can divide the contribution to the superconducting order parameter into the on-site, first-neighbor, second-neighbor contributions, etc. Here, we will construct the on-site and first-neighbor contribution to the superconducting order parameter.

\subsection{On-site approximation}\label{onsite}
In the case of the on-site interaction, only the diagonal matrix elements of the gap function of the type
$\Delta^{\rm AA/BB}$ are nonzero and {\bf k}-independent. For the spin channels $d_{z}$ and $(d_x,d_y)$ it is impossible to satisfy the Pauli principle. On the other hand, $\Delta^{A\sigma,A\sigma'}= -\Delta^{A\sigma',A\sigma}$ (and equivalently $\Delta^{B\sigma,B\sigma'}= -\Delta^{B\sigma',B\sigma}$) is satisfied in the singlet case~\cite{Pangburn2023}.
For the singlet spin channel, the superconducting gap function can be obtained using the following construction
\begin{eqnarray}
\Delta^{\rm AA/BB}_{A_m,(s)}&=&\Delta_{\rm A/B}\sum_{g\in {\bf C}_{3}}\Gamma_{A_m}^*(g)d_0,
\end{eqnarray}
where $g=C_3^u$, $u=0,1,2$ are the elements of the ${\bf C}_3$ group equal to the identity element ($u=0$) and rotation around the z-axis for an angle of $2\pi/3u$ ($u=1,2$),
while $\Gamma_{A_m}^*(g)$ is the conjugated matrix element of the IR $A_m$ for the given group element $g$, see Table~\ref{tab:C3v}.
Finally, we note that $\Delta_{\rm A}$ and $\Delta_{\rm B}$ represent superconducting pairing strengths on sublattices A and B, respectively, that do not have to be equal, since there is no group element that connects carbon atoms on sublattice A and B.
Simple algebra shows that the only nonzero $\Delta^{\rm AA/BB}_{A_m,(s)}$ blocks are those transforming according to the IR $A_0$ and are equal to $\Delta^{\rm AA/BB}_{A_0,(s)}=\Delta_{\rm A/B}d_{0}$.
This gives us a gap function in the on-site approximation
\begin{eqnarray}\label{onsite}
\Delta_{A_0, (s)}^{\rm os}(\phi)=\begin{pmatrix}
    0 & \cos{\phi} &0 &0 \\
    -\cos{\phi} & 0 &0 &0\\
    0 &  0 &0 & \sin{\phi}\\
   0 & 0 &-\sin{\phi} & 0\\
\end{pmatrix}.
\end{eqnarray}
In the previous equation, we defined a dimensionless parameter $\phi\in(0,2\pi)$ such that $\Delta_{\rm A}=\cos{\phi}$ and $\Delta_{\rm B}=\sin{\phi}$ represent normalized gap intensity strengths that can be used to analyze the imbalance between the sublattice contributions.

In the second-quantized form, 
gap operator $\hat{\Delta}_{A_0, (s)}^{\rm os}=\sum_{{\bf k}}\psi_{\bf k}^{\dagger}\Delta_{A_0, (s)}^{\rm os}(\phi)\varphi_{\bf k}$, where $\psi_{\bf k}^{\dagger}=[a_{\bf k\uparrow}^{\dagger},
a_{\bf k\downarrow}^{\dagger},b_{\bf k\uparrow}^{\dagger},b_{\bf k\downarrow}^{\dagger}]$, $\varphi_{\bf k}=[a_{-\bf k\uparrow}^{\dagger},
a_{-\bf k\downarrow}^{\dagger},b_{\bf -k\uparrow}^{\dagger},b_{\bf -k\downarrow}^{\dagger}]^{\rm T}$, is equal to
\begin{eqnarray}
\hat{\Delta}_{A_0, (s)}^{\rm os}&=&\sum_{{\bf k}}\cos{\phi} 
(a_{{\bf k}\uparrow}^{\dagger}a_{-{\bf k}\downarrow}^{\dagger}-
a_{{\bf k}\downarrow}^{\dagger}a_{-{\bf k}\uparrow}^{\dagger})\nonumber\\
&&+\sin{\phi} (b_{{\bf k}\uparrow}^{\dagger}b_{-{\bf k}\downarrow}^{\dagger}-
b_{{\bf k}\downarrow}^{\dagger}b_{-{\bf k}\uparrow}^{\dagger}).
\end{eqnarray}

\begin{table}[t]
\setlength{\tabcolsep}{10pt}
\renewcommand{\arraystretch}{1.5}
\caption{Character table of IRs $A_0$, $A_1$, and $A_{-1}$ of the 
${\bf C}_{3}$ group, where $C_3^u$ ($u=0,1,2$) represents 
rotation for an angle of $2\pi u/3$ around the $z$-axis, 
perpendicular to the graphene plane. The complex characters of IRs $A_{\pm 1}$ reflect the fact that these IRs carry nonzero angular momenta ($m = \pm 1$ to be fully precise), representing chiral objects that spontaneously break TRS. Thus, gap functions transforming according to $A_{\pm 1}$ describe chiral superconducting 
states of opposite chirality, forming time-reversal pairs. On the other hand, $A_0$ is the trivial real IR carrying zero angular momentum, and gap functions transforming according to $A_0$ must preserve TRS.}
\label{tab:C3v}
\centering
\begin{tabular}{|c|c|c|c|}
  IR  & $C_3^0$ & $C_3^1$ & $C_3^2$ \\ 
  $A_0$ & 1        & 1 &    1     \\ 
  $A_1$ & 1& ${\rm e}^{{\rm i}\frac{2\pi}3}$    &  ${\rm e}^{-{\rm i}\frac{2\pi}3}$       \\ 
 $A_{-1}$ & 1& ${\rm e}^{-{\rm i}\frac{2\pi}3}$   &  ${\rm e}^{{\rm i}\frac{2\pi}3}$       \\    
\end{tabular}
\end{table}

\subsection{Nearest-neighbor approximation}\label{nnAPPROX}
In the the nearest-neighbor approximation, one has to consider the hopping between the atoms on sublattice A and the atoms on sublattice B, having the distance $a/\sqrt{3}$ between themselves, where $a$ is the lattice constant of graphene. This implies zero $\Delta^{\rm AA/BB}_{\bf k}$  blocks and non-zero $\Delta^{\rm AB/BA}_{\bf k}$ blocks. 
\subsubsection{Singlet $d_0$ gap functions}
In the nearest-neighbor approximation, the $\Delta^{\rm AB}_{{\bf k},A_m(s)}$ block in the singlet case can be constructed as
\begin{eqnarray}
\Delta^{\rm AB}_{{\bf k},A_m(s)}&=&\sum_{g\in {\bf C}_{3}}\Gamma_{A_m}^*(g)
{\rm e}^{{\rm i}{\bf k}\cdot(\mathcal{D}(g^{-1}){\bf r}_{AB'})}d_0,
\end{eqnarray}
where $m=0,\pm 1$, while ${\bf r}_{\rm nn}=a/\sqrt{3}(0,1,0)$ represents the distance between the representative nearest-neghbor atoms on sublattice A and B (see FIG.~\ref{FIG1}c), while $\mathcal{D}$ corresponds to the matrix representation transforming normal and pseudovectors in the same way
\begin{eqnarray}
\mathcal{D}(C_3^u)=\begin{pmatrix}
    \cos{\frac{2\pi}3u} & -\sin{\frac{2\pi}3u} & 0  \\
    \sin{\frac{2\pi}3u} & \cos{\frac{2\pi}3u} & 0   \\
      0 & 0 & 1
\end{pmatrix}.
\end{eqnarray}
After a simple algebra, we get $\Delta^{\rm AB}_{{\bf k},A_m(s)}=f_{m}({\bf k})d_0$,
where $f_m(\bm k)=e^{{\rm i}\frac{a k_y}{\sqrt{3}}}+2{\rm e}^{-{\rm i}\frac{ak_y}{2\sqrt{3}}}\cos{(\frac{a k_x}2-m\frac{2\pi}3)}$. Thus, $\Delta_{A_m, (s)}^{\rm nn}({\bf k})$ is equal to
\begin{eqnarray}
\Delta_{A_m, (s)}^{\rm nn}({\bf k})&=&
\begin{pmatrix}
\begin{smallmatrix}
    0 & 0 & 0 &f_m({\bf k}) \\
    0 & 0 & -f_m({\bf k}) &0\\
    0 & f_m(-{\bf k}) & 0 & 0\\
   -f_m(-{\bf k}) & 0 &0 & 0\\
\end{smallmatrix}
\end{pmatrix},
\end{eqnarray}
while in the second quantized form, $\hat{\Delta}_{A_m, (s)}^{\rm nn}$ is equal to
\begin{eqnarray}
\hat{\Delta}_{A_m, (s)}^{\rm nn}&=&\sum_{\bf k}\Big[f_m({\bf k})(
a_{{\bf k}\uparrow}^{\dagger}b_{-{\bf k}\downarrow}^{\dagger}-
a_{{\bf k}\downarrow}^{\dagger}b_{-{\bf k}\uparrow}^{\dagger})\nonumber\\
&&+f_m(-{\bf k})(
b_{{\bf k}\uparrow}^{\dagger}a_{-{\bf k}\downarrow}^{\dagger}-
b_{{\bf k}\downarrow}^{\dagger}a_{-{\bf k}\uparrow}^{\dagger}
)\Big].
\end{eqnarray}
An important check of the validity of our construction is the behavior of the constructed gap functions with respect to the TRS. The TRS operator $\Theta$, in the second quantization picture, has the following action on the components of the gap functions $\hat{\Delta}_{A_m, (s)}^{\rm nn}$:
$\Theta \chi_{\bf k\uparrow}^{\dagger}\Theta^{-1}=\chi_{-\bf k\downarrow}^{\dagger}$, $\Theta \chi_{\bf k\downarrow}^{\dagger}\Theta^{-1}=-\chi_{-\bf k\uparrow}^{\dagger}$, $\chi=a,b$, while $\Theta f\Theta^{-1}=f^*$, where f is an arbitrary scalar. The gap function $\hat{\Delta}_{A_m, (s)}^{\rm nn}$ satisfies the time-reversal symmetry if the relation $\Theta\hat{\Delta}_{A_m, (s)}^{\rm nn}\Theta^{-1}=\hat{\Delta}_{A_m, (s)}^{\rm nn}$ is fulfilled. After calculating
\begin{eqnarray}
    \Theta\hat{\Delta}_{A_m, (s)}^{\rm nn}\Theta^{-1}&=&\sum_{\bf k}\Big[f_m^*(-{\bf k})(
a_{{\bf k}\uparrow}^{\dagger}b_{-{\bf k}\downarrow}^{\dagger}-
a_{{\bf k}\downarrow}^{\dagger}b_{-{\bf k}\uparrow}^{\dagger})\nonumber\\
&&+f_m^*({\bf k})(
b_{{\bf k}\uparrow}^{\dagger}a_{-{\bf k}\downarrow}^{\dagger}-
b_{{\bf k}\downarrow}^{\dagger}a_{-{\bf k}\uparrow}^{\dagger}
)\Big],
\end{eqnarray}
we confirm that the TRS is satisfied if the relation $f_m(k)=f_m^*(-k)$ holds, which is the case only for IR $A_0$ ($\hat{\Delta}_{A_0, (s)}^{\rm nn}$). We mention that $\hat{\Delta}_{A_1, (s)}^{\rm nn}$ and $\hat{\Delta}_{A_{-1}, (s)}^{\rm nn}$ represent time-reversal pairs, since the relation $\Theta\hat{\Delta}_{A_1, (s)}^{\rm nn}\Theta^{-1}=\hat{\Delta}_{A_{-1}, (s)}^{\rm nn}$ is satisfied, which is consistent with the nature of IRs $A_1$ and $A_{-1}$, representing time-reversal pairs. 

\subsubsection{Triplet $d_z$ gap functions}\label{dzsub}
In the case of the $d_z$ triplet gap functions, the same construction as in the singlet case, i.e. using the formula $\sum_{g\in {\bf C}_{3}}\Gamma_{A_m}^*(g)
{\rm e}^{{\rm i}{\bf k}\cdot(\mathcal{D}(g^{-1}){\bf r}_{AB'})}\mathcal{D}(g^{-1})d_z=f_{m}({\bf k})d_z$, does not give us gap function that have the proper behavior with respect to the TRS, $\Theta\hat{\Delta}_{A_0, (z)}^{\rm nn}\Theta^{-1}=\hat{\Delta}_{A_0, (z)}^{\rm nn}$ and 
$\Theta\hat{\Delta}_{A_1, (z)}^{\rm nn}\Theta^{-1}=\hat{\Delta}_{A_{-1}, (z)}^{\rm nn}$. However, a proper behavior of the gap functions is obtained using the alternative definition
\begin{eqnarray}
\Delta^{\rm AB}_{{\bf k},A_m(z)}&=&\sum_{g\in {\bf C}_{3}}\Gamma_{A_m}^*(g)
\sin{({\bf k}\cdot(\mathcal{D}(g^{-1}){\bf r}_{AB'})}\mathcal{D}(g^{-1})d_z\nonumber\\
&=& g_m({\bf k})d_z,
\end{eqnarray}
where $g_0({\bf k})=\sin{\frac{a k_y}{\sqrt{3}}}-2\cos{\frac{a k_x}2}\sin{\frac{a k_y}{2\sqrt{3}}}$, while $g_{\pm 1}({\bf k})$
are equal to $g_{\pm 1}({\bf k})=\sin{\frac{a k_y}{\sqrt{3}}}+\cos{\frac{a k_x}{2}}\sin{\frac{a k_y}{2\sqrt{3}}}\mp{\rm i}\sqrt{3}\sin{\frac{a k_x}{2}}\cos{\frac{a k_y}{2\sqrt{3}}}$. Thus, $\Delta_{A_m, (z)}^{\rm nn}({\bf k})$ is equal to
\begin{eqnarray}
\Delta_{A_m, (z)}^{\rm nn}({\bf k})&=&
\begin{pmatrix}
\begin{smallmatrix}
    0 & 0 & 0 &g_m({\bf k}) \\
    0 & 0 & g_m({\bf k}) &0\\
    0 & -g_m(-{\bf k}) & 0 & 0\\
   -g_m(-{\bf k}) & 0 &0 & 0\\
\end{smallmatrix}
\end{pmatrix},
\end{eqnarray}
or in the second quantized form,
\begin{eqnarray}\label{tripletZ}
\hat{\Delta}_{A_m, (z)}^{\rm nn}&=&\sum_{\bf k}\Big[g_m({\bf k})(
a_{{\bf k}\uparrow}^{\dagger}b_{-{\bf k}\downarrow}^{\dagger}+
a_{{\bf k}\downarrow}^{\dagger}b_{-{\bf k}\uparrow}^{\dagger})\nonumber\\
&&-g_m(-{\bf k})(
b_{{\bf k}\uparrow}^{\dagger}a_{-{\bf k}\downarrow}^{\dagger}+
b_{{\bf k}\downarrow}^{\dagger}a_{-{\bf k}\uparrow}^{\dagger}
)\Big].
\end{eqnarray}
The gap functions constructed using the Eq.~\eqref{tripletZ} 
have the correct TRS behavior, since $\hat{\Delta}_{A_0, (z)}^{\rm nn}$ 
preserves TRS, while $\hat{\Delta}_{A_1, (z)}^{\rm nn}$ and 
$\hat{\Delta}_{A_{-1}, (z)}^{\rm nn}$ are time-reversal partners.

\subsubsection{Triplet $(d_x,d_y)$ gap functions}

Finally, we construct the gap functions in the $(d_x,d_y)$ space, and demand, similarly as in the previous two cases, the proper TRS behavior of the constructed gap functions, i.e. $\Theta\hat{\Delta}_{A_0, (xy)}^{\rm nn}\Theta^{-1}=\hat{\Delta}_{A_0, (xy)}^{\rm nn}$ and 
$\Theta\hat{\Delta}_{A_1, (xy)}^{\rm nn}\Theta^{-1}=\hat{\Delta}_{A_{-1}, (xy)}^{\rm nn}$. Such a behavior can be obtained by defining a gap function in the following manner
\begin{eqnarray}
\Delta^{\rm AB}_{{\bf k},A_m(xy)}&=&\sum_{g\in {\bf C}_{3}}\Gamma_{A_m}^*(g)
\sin{({\bf k}\cdot(\mathcal{D}(g^{-1}){\bf r}_{AB'})}\mathcal{D}(g^{-1}){\bf d}_1,\nonumber\\
&=&h_m^x({\bf k})d_x+h_m^y({\bf k})d_y,
\end{eqnarray}
where the functions $h_{m}^x({\bf k})$ and 
$h_{m}^y({\bf k})$ are equal to
\begin{eqnarray}
    h_0^x({\bf k})&=&\cos{\frac{a k_x}2}\sin{\frac{a k_y}{2\sqrt{3}}}+\sin{\frac{a k_y}{\sqrt{3}}},\\
    h_0^y({\bf k})&=&-\sqrt{3}\sin{\frac{a k_x}2}\cos{\frac{a k_y}{2\sqrt{3}}},\nonumber\\
     h_{\pm 1}^x({\bf k})&=&-\frac{1}2\cos{\frac{a k_x}2}\sin{\frac{a k_y}{2\sqrt{3}}}+\sin{\frac{a k_y}{\sqrt{3}}}\nonumber\\
     &&\pm{\rm i}\frac{\sqrt{3}}{2}\sin{\frac{a k_x}2}\cos{\frac{a k_y}{2\sqrt{3}}},\nonumber\\
    h_{\pm 1}^y({\bf k})&=&\frac{\sqrt{3}}2\sin{\frac{a k_x}2}\cos{\frac{a k_y}{2\sqrt{3}}}
     \mp{\rm i}\frac{3}{2}\cos{\frac{a k_x}2}\sin{\frac{a k_y}{2\sqrt{3}}}.\nonumber
\end{eqnarray}
Taking this into the account, we get $\Delta_{A_m, (xy)}^{\rm nn}({\bf k})$, 
\begin{eqnarray}
\Delta_{A_m, (xy)}^{\rm nn}({\bf k})&=&
\begin{pmatrix}
\begin{smallmatrix}
    0 & 0 & -h_m^x({\bf k}) &0 \\
    0 & 0 & 0 & h_m^x({\bf k})\\
    h_m^x(-{\bf k}) &0 & 0 & 0\\
   0 & -h_m^x(-{\bf k}) &0 & 0\\
   \end{smallmatrix}
\end{pmatrix}+\nonumber\\
&&\begin{pmatrix}
\begin{smallmatrix}
    0 & 0 & {\rm i} h_m^y({\bf k}) &0 \\
    0 & 0 & 0 & {\rm i} h_m^y({\bf k})\\
    -{\rm i} h_m^y(-{\bf k}) &0 & 0 & 0\\
   0 & -{\rm i} h_m^y(-{\bf k}) &0 & 0\\
   \end{smallmatrix}
\end{pmatrix}.\nonumber\\
\end{eqnarray}
Finally, in the second quantized form, 
gap operator $\hat{\Delta}_{A_m, (xy)}^{\rm nn}$ in the nearest-neighbor approximation is equal to
\begin{eqnarray}
\hat{\Delta}_{A_m, (xy)}^{\rm nn}&=&\sum_{\bf k}\Big[h_m^x({\bf k}) (-a_{{\bf k}\uparrow}^{\dagger}b_{-{\bf k}\uparrow}^{\dagger}+
a_{{\bf k}\downarrow}^{\dagger}b_{-{\bf k}\downarrow}^{\dagger})+\nonumber\\
&&h_m^x(-{\bf k})(
b_{{\bf k}\uparrow}^{\dagger}a_{-{\bf k}\uparrow}^{\dagger}-
b_{{\bf k}\downarrow}^{\dagger}a_{-{\bf k}\downarrow}^{\dagger}
)+\nonumber\\
&&{\rm i}h_m^y({\bf k})(a_{{\bf k}\uparrow}^{\dagger}b_{-{\bf k}\uparrow}^{\dagger}+
a_{{\bf k}\downarrow}^{\dagger}b_{-{\bf k}\downarrow}^{\dagger}))+\nonumber\\
&&-{\rm i}h_m^y(-{\bf k})(
b_{{\bf k}\uparrow}^{\dagger}a_{-{\bf k}\uparrow}^{\dagger}+
b_{{\bf k}\downarrow}^{\dagger}a_{-{\bf k}\downarrow}^{\dagger})\Big].
\end{eqnarray}

\subsubsection{Singlet-triplet mixing}
The classification of the superconducting gap functions in terms of the IRs of the ${\bf C}_3$ group allows us to mix all gap functions belonging to the same IR. This is because gap functions within the same IR share identical transformation properties under all symmetry operations of the group, making them indistinguishable from a symmetry 
perspective. In the case of the IR $A_0$, the general form of the superconducting gap function $\Delta_{A_0}({\bf k})$ mixes the on-site singlet contribution $\Delta_{A_0, (s)}^{\rm os}(\phi)$, nearest-neighbor singlet contribution $\Delta_{A_0, (s)}^{\rm nn}({\bf k})$, and two triplet components $\Delta_{A_0, (z)}^{\rm nn}({\bf k})$, and $\Delta_{A_0, (xy)}^{\rm nn}({\bf k})$. 

On the other hand, using the derived superconducting pairings that transform according to the IRs $A_{\pm 1}$, we can define a singlet-triplet mixing gap function
\begin{eqnarray}\label{Apm1}
    \Delta_{A_{\pm 1}}({\bf k})&=&\Delta_0 \Big[\cos{\theta}\Delta_{A_{\pm 1}, (s)}^{\rm nn}({\bf k})+
    \sin{\theta}\cos{\phi_t}\Delta_{A_{\pm 1}, (z)}^{\rm nn}({\bf k})\nonumber\\
    &&+\sin{\theta}\sin{\phi_t}\Delta_{A_{\pm 1}, (xy)}^{\rm nn}({\bf k})\Big],
\end{eqnarray}
where the parameter $\theta$ determines the level of the singlet-triplet mixing, $\phi_t$ quantifies the mixing between triplet components $\Delta_{A_{\pm 1}, (z)}^{\rm nn}({\bf k})$ and $\Delta_{A_{\pm 1}, (xy)}^{\rm nn}({\bf k})$, while  $\Delta_0$ represents the gap intensity, set in our calculation to 1~{\rm meV}. 

This value of $\Delta_0$ represents an upper bound for the superconducting gap 
induced in graphene by proximity to NbSe$_2$. Bulk NbSe$_2$ is a 
superconductor whose measured intrinsic gap is  $\sim2.6$~meV~\cite{Sanchez1995}. In the monolayer limit, where inversion 
symmetry is broken and Ising superconductivity is dominant~\cite{Xi2016}, 
this gap is further reduced~\cite{Khestanova2018}. The proximity-induced 
gap transferred to graphene is expected to be even smaller. Measurements in NbSe$_2$-based van der Waals heterostructures suggest 
induced gaps of only $\sim0.1$-$0.2$~meV at the surface of the 
proximitized layer~\cite{Dai2017}. Our choice of $\Delta_0 = 1$~meV 
therefore sets an upper bound.
 At the same time, DFT calculations on twisted graphene/NbSe$_2$ heterostructures show
that the proximity-induced Rashba SOC reaches values $0.225-2.638$~meV~\cite{NGF24}, depending on the twist angle,
suggesting that the assumed $\lambda_R \geq \Delta_0$ regime is fully justified.

\section{Topological implications}\label{topology}

Depending on the presence/absence of the TRS in the system, topological phases can be classified using the $\mathbb{Z}_2$ topological invariant (TRS case) or the Chern number $C$ (broken TRS). Since the studied graphene/NbSe$_2$ heterostructure is nonmagnetic, TRS is preserved in the normal phase (thus for both electrons and holes).
In the superconducting phase, TRS is conserved/broken if the gap function transforms according to the IR $A_0$/$A_{\pm 1}$. Therefore, we use the $\mathbb{Z}_2$ invariant to classify the topological phases of the BdG Hamiltonian with gap function  $\Delta_{A_0}({\bf k})$, which takes the value $\mathbb{Z}_2 = 0$ (trivial) or $\mathbb{Z}_2 = 1$ (topological). Oppositely, $C$ is used to classify the topological phases in the case when the  BdG Hamiltonian is constructed using the $\Delta_{A_{\pm 1}}(\mathbf{k})$~\eqref{Apm1} gap function, where $C = 0$ indicates a trivial state, while $C \neq 0$ suggests a topologically 
nontrivial phase~\cite{asboth2016short,maciejko2010topological,wieder2021topological,HK10}. 

\begin{figure}[t]
\centering
\includegraphics[width=0.49\textwidth]{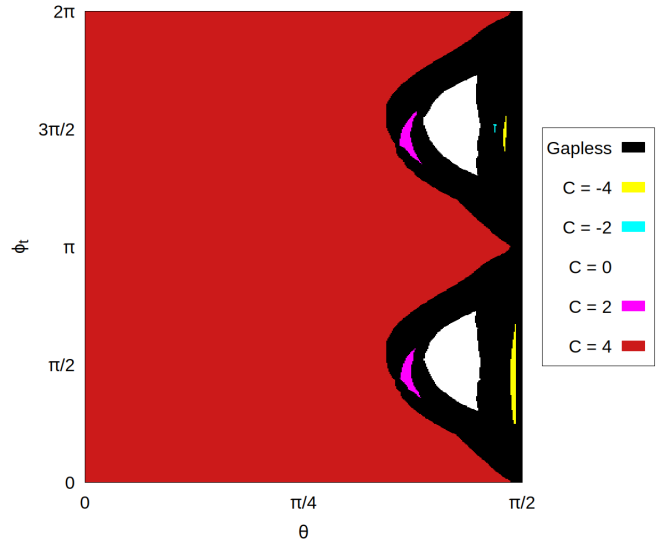}
\caption{Chern number $C$ phase diagram of the BdG Hamiltonian with the gap function $\Delta_{A_1}(\mathbf{k})$, 
Eq.~\eqref{Apm1}, as a function of the singlet-triplet mixing angle 
$\theta \in (0, \pi/2)$ and the inter-triplet mixing angle 
$\phi_t \in (0, 2\pi)$. Different colors correspond to different
$C$ phases: trivial phase $C = 0$ (white), 
topological phases with $C = \pm 2$ (cyan/magenta) and $C = \pm 4$ 
(yellow/dark red), and gapless regions (black).}\label{ChernFIGURE}
\end{figure}

Topological invariants were obtained through Wilson-loop 
calculations~\cite{Yu2011}, following the approach of Fu and 
Kane~\cite{Fu2006}. A detailed explanation of the method and the numerical details can be found in 
Appendix~\ref{app:wilson_loop}. 
To classify the topological phases in the system, one has to first
determine whether a global band gap is present, since only in the case of a fully gapped spectrum 
topological invariant  are well defined and can be extracted from the occupied bands. 
A main numerical challenge when calculating both invariants is in the accurate calculation 
of the BdG gap across different $k$-points in the Brillouin zone (BZ). In graphene-based systems, the relevant 
low-energy physics is in the vicinity of the $K$ and $K'$ points of graphene. This is precisely
where the BdG gap opens (at the normal-state Fermi-level contours 
around these valleys) and where the WFC flow exhibits its 
nontrivial behavior. The size of the region around K (K') point in reciprocal space is dependent on the chemical potential $\mu$. For $\mu \approx 0$ only a small neighborhood 
of $K$ and $K'$ is relevant, whereas at the large doping as in our system 
($\mu \approx 525$~meV) this region is much bigger and requires a 
dense $\mathbf{k}$-point mesh to resolve both the band gap and the topological invariant. 

Our results indicate that $\mathbb{Z}_2 = 0$ for BdG Hamiltonian including gap functions transforming according to IR $A_0$, throughout the sampled 
parameter space. More precisely, $\mathbb{Z}_2 = 0$ for both the purely singlet gap function 
$\Delta^s_{A_0}(\theta, \varphi)$, which mixes on-site and nearest-neighbor 
singlet components, and the singlet-triplet mixed gap function 
$\Delta_{A_0}(\theta, \phi_t)$,  which additionally incorporates triplet 
pairing channels (for more details see Appendix~\ref{Z2app}). We mention that the determination of $\mathbb{Z}_2$ topological invariant is numerically subtle, since the near-degeneracy of the WFC flow due to small SOC resembles a topological $\mathbb{Z}_2 = 1$ phase at insufficient $\mathbf{k}$-point 
mesh. Due to that, a careful analysis with increased sampling is required to 
properly identify $\mathbb{Z}_2 = 0$, as illustrated in 
FIG.~\ref{WLsingletAO} and discussed in detail in 
Appendix~\ref{app:wilson_loop}. 

In contrast to the $\Delta_{A_0}({\bf k})$ case, the presence of the TRS-breaking $\Delta_{A_{\pm 1}}({\bf k})$ gap function triggers a rich topological phase diagram, as shown in FIG.~\ref{ChernFIGURE}. Since the IRs $A_1$ and $A_{-1}$ are related by TRS, in the sense that applying the time-reversal operator maps 
$\Delta_{A_1}(\mathbf{k})$ into $\Delta_{A_{-1}}(\mathbf{k})$, the Chern numbers satisfy $C_{A_1} = -C_{A_{-1}}$, and it is enough to analyze the single IR, for which we choose $A_1$. The gap function 
mixes all three nearest-neighbor pairing channels present in the $A_1$ IR, with $\theta$ quantifying the level of singlet-triplet mixing and $\phi_t$ mixing between two triplet components 
$\Delta_{A_1,(z)}^{\rm nn}(\mathbf{k})$ and $\Delta_{A_1,(xy)}^{\rm nn}(\mathbf{k})$, see~\eqref{Apm1}. Since the gap function satisfies
$\Delta_{A_1}(\theta + \pi, \phi_t) = -\Delta_{A_1}(\theta, \phi_t)$ and
$\Delta_{A_1}(\pi - \theta, \phi_t + \pi) = -\Delta_{A_1}(\theta, \phi_t)$,
and an overall gap sign change leaves all physical observables invariant, it suffices to restrict without loss of generality to $\theta \in (0, \pi/2)$,
$\phi_t \in (0, 2\pi)$. The full phase diagram over $\theta \in (0, 2\pi)$,
$\phi_t \in (0, 2\pi)$ can then be reconstructed using the resulting symmetry
relations for the Chern number, $C(\theta + \pi, \phi_t) = C(\theta, \phi_t)$
and $C(\pi - \theta, \phi_t + \pi) = C(\theta, \phi_t)$.

The phase diagram of FIG.~\ref{ChernFIGURE} was obtained using a two-step 
procedure due to the need to balance computational cost and the need to resolve the BdG gap and WFC flow near $K$ and $K'$ accurately. In the first step, the full parameter 
space $\theta \in (0, \pi/2)$, $\phi_t \in (0, 2\pi)$ was sampled on a coarse 
$48 \times 192$ $(\theta, \phi_t)$ grid, with topological invariants and gap 
values evaluated on a uniform $6000 \times 6000$ $\mathbf{k}$-point mesh. This step reveals that for $\theta \in (0, 
\pi/3)$ the Chern number takes the value $C = 4$ independently of $\phi_t$, 
with a minimum gap value of $0.136$~meV, confirming that this region is fully gapped 
and topologically nontrivial throughout. In the second step, the remaining region $\theta \in (\pi/3, \pi/2)$, $\phi_t \in (0, 2\pi)$, where gap closings and topological phase transitions occur, was sampled on a refined $128 \times 768$ $(\theta, \phi_t)$ grid. To accurately resolve the gap near $K$ and $K'$, gap values were computed on a uniform $18000 \times 18000$ $\mathbf{k}$-point mesh. At the same time, the WFC flow was evaluated on two strips of $1500$ points along $\mathbf{b}_1$, centered on $\tfrac{1}{3}\mathbf{b}_1$ and $\tfrac{2}{3}\mathbf{b}_1$, combined with a full $18000$-point mesh along $\mathbf{b}_2$ (see FIG.~\ref{WLtriplet} and Appendix~\ref{ChernNUMBER} for full details). Configurations with a gap smaller than $0.04$~meV were classified as gapless states. 

To verify and interpret the nature of the phase transitions, we 
computed the local Chern number contribution using the 
Fukui-Hatsugai-Suzuki (FHS) method~\cite{FHS05}. The FHS method 
defines the plaquette lattice curvature via the closed 
parallel-transport operator
\begin{equation}
\widetilde{F}_{12}(\mathbf{k})=\log\!\left[
U_1(\mathbf{k})U_2(\mathbf{k}+\delta\mathbf{k}_1)
U_1^{-1}(\mathbf{k}+\delta\mathbf{k}_2)U_2^{-1}(\mathbf{k})
    \right],
\end{equation}
where the overlap matrix and the corresponding $U(1)$ link 
variable are defined as $\mathcal{U}_{\mu}^{mn}(\mathbf{k})
    =
    \langle u_m(\mathbf{k})|u_n(\mathbf{k}+\delta\mathbf{k}_{\mu})\rangle$,
    leading to $U_{\mu}(\mathbf{k})
    =
    \det \mathcal{U}_{\mu}(\mathbf{k})/
    |\det \mathcal{U}_{\mu}(\mathbf{k})|$.
The overlap matrix $\mathcal{U}_\mu(\mathbf{k})$ coincides with 
the elementary overlap matrix $M^{i,i+1}(k_1)$ defined in 
Appendix~\ref{app:wilson_loop}, while $U_\mu(\mathbf{k})$ is 
obtained by extracting the phase of the 
determinant. This construction is well known within the lattice 
gauge theory formalism~\cite{Tong}. For a sufficiently small 
plaquette, the link variables $U_\mu$ may be viewed as short 
parallel transports generated by the Berry connection, and the 
plaquette operator $\widetilde{F}_{12}(\mathbf{k})$ measures 
the Berry curvature enclosed by the small loop. The local Chern 
number contribution of each plaquette is then 
$q_{12}(\mathbf{k}) = (1/2\pi)\,\mathrm{Im}\,
\widetilde{F}_{12}(\mathbf{k})$, and the total Chern number can be calculated by summing over all plaquettes, 
$C = \sum_{\mathbf{k}} q_{12}(\mathbf{k})$.
\begin{figure*}[t]
\centering
\includegraphics[width=0.87\textwidth]{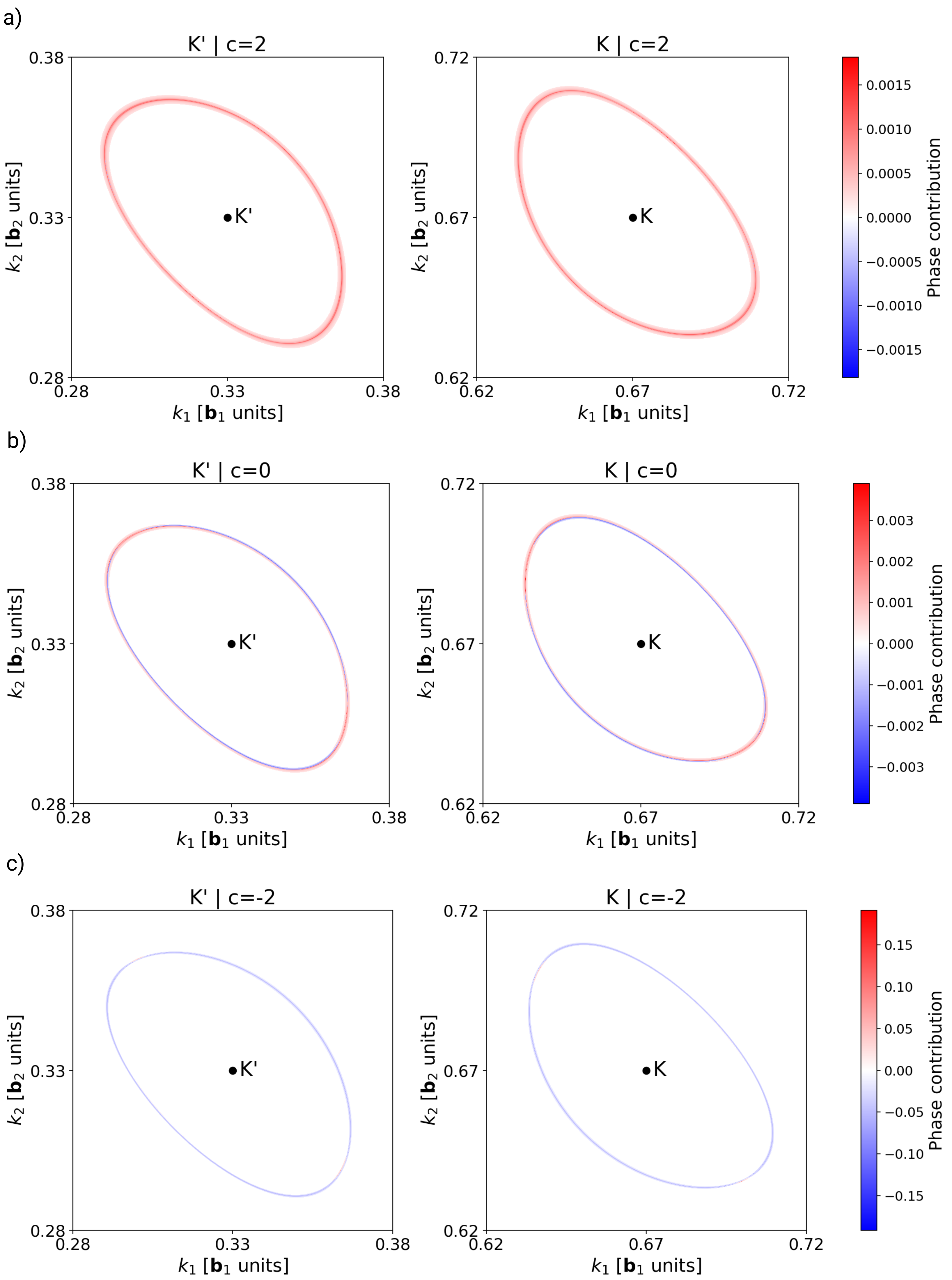}
\caption{Local Chern number contributions  in small regions around the $K$ and $K'$ valleys for three representative 
points in the phase diagram of FIG.~\eqref{ChernFIGURE} at fixed $\phi_t = 1.55$. (a) $C=4$ phase ($\theta = 1.00$). The plaquette 
contribution is sharply focused on a closed loop encircling each 
valley, with $C_K = C_{K'} = 2$, giving a total 
Chern number $C = 4$. 
(b) $C=0$ phase ($\theta = 1.35$). The contribution remains 
localized around the valley contours, but positive and negative 
parts around each valley cancel, yielding 
$C_K = C_{K'} = 0$ and $C = 0$. (c) $C=-4$ phase ($\theta = 1.58$). The Berry-curvature weight remains localized near the two valleys but its sign is 
reversed relative to (a), giving $C_K = C_{K'} = -2$ and $C = -4$.}\label{FHZ}
\end{figure*} 

The FHS method-based calculations reveal that the Berry curvature is  concentrated in small regions around the two graphene 
valleys, $K$ and $K'$. This is expected, since the 
superconducting gap is on the meV scale and affects the BdG 
eigenvectors only in a narrow momentum-space window around the 
Fermi-level contours encircling $K$ and $K'$ (the size of 
these contours is set by the chemical potential, as discussed 
above) leaving the rest of the BZ essentially unaffected. 
After evaluating $q_{12}(\mathbf{k})$ on two valley patches 
$\mathcal{R}_K$ and $\mathcal{R}_{K'}$ enclosing gapped valley contours, the sum of the valley contributions 
$C = C(\mathcal{R}_K) + C(\mathcal{R}_{K'})$ 
reproduces the full Wilson-loop Chern number to a precision 
better than $10^{-5}$. We focus of three examples, $C \in \{4, 0, -4\}$, and show that the Chern number contribution entirely comes from the two valley regions, as illustrated 
in FIG.~\ref{FHZ}. In the $C=4$ phase (FIG.~\ref{FHZ}a), 
the plaquette contribution is sharply focused on closed loops 
around each valley, with $C_K = C_{K'} = 2$. In the trivial 
phase (FIG.~\ref{FHZ}b), positive and negative contributions 
cancel within each valley patch, giving $C_K = C_{K'} = 0$. Finally, in the $C=-4$ phase (FIG.~\ref{FHZ}c), the Berry-curvature weights are reversed with respect to (a) case, since 
$C_K = C_{K'} = -2$.

An important question concerns the physical realizability of the 
topologically nontrivial phases found within IRs $A_{\pm 1}$. To address 
this, it is useful to connect our results to the symmetry analysis 
of the superconducting gap functions in monolayer NbSe$_2$ itself. In the absence of a twist, monolayer NbSe$_2$ possesses the full ${\bf D}_{3{\rm h}}$ point group symmetry, and the superconducting order parameters can be classified according to its IRs~\cite{Milivojevic2026}. Within this classification, the gap functions transforming according to the IR $E^{g/u}$ can have different realizations: a nematic phase, which preserves TRS, and a chiral phase, conventionally written as $(1, i)$ 
or $(1, -i)$, which spontaneously breaks it~\cite{Hanis2024}. The nature of the superconducting order parameter in monolayer NbSe$_2$ is actively debated, with proposals ranging from dominant conventional $s$-wave pairing~\cite{Engstrom2025}, to chiral superconducting phase~\cite{Siegl2025}, and nodal or nematic order arising from competing superconducting channels~\cite{CLA+22}. 

The formation of the van der Waals heterostructure and the symmetry reduction can have a profound effect on both materials. Using the group theory terminology, the twist-induced 
reduction of the point group symmetry from ${\bf D}_{3{\rm h}}$ to 
${\bf C}_3$ splits the degenerate two-dimensional $E^{g/u}$ 
IR into two nondegenerate one-dimensional IRs, precisely the chiral IRs $A_1$ and $A_{-1}$ of ${\bf C}_3$. Thus, the symmetry reduction ${\bf D}_{3{\rm h}} \to {\bf C}_3$  naturally favors the chiral $A_1$ or $A_{-1}$ states over the nematic one. In other words, the twist angle acts as a symmetry-based chirality selector. 

On the graphene side, the superconducting proximity effect is triggered, accompanied with the 
spin-orbit proximity effect that induces sizable intrinsic and Rashba SOC, as well as the nonzero Rashba phase $\phi_R$ that is a consequence of the nonzero twist 
angle. On the NbSe$_2$ side, the reduced symmetry may shift the relative stability of competing superconducting channels, potentially promoting a subdominant chiral component to become dominant. The symmetry type of superconducting order parameter stabilized in 
NbSe$_2$ directly determines the character of the superconducting
pairing proximity-induced into graphene. If a chiral superconducting phase transforming according to IRs $A_1$ or 
$A_{-1}$ is stabilized in NbSe$_2$, the proximity-induced superconductivity in graphene will acquire the same 
TRS-breaking character.

Lastly, we briefly comment on two possible experimental signatures of the TRS-breaking chiral phases. First, it has recently been shown that chiral superconductivity leaves a distinctive quasiparticle
interference fingerprint at atomic defects~\cite{Chiral_QPI2025}, 
promoting quasiparticle interference imaging as a tool for identifying TRS-breaking 
pairing in two-dimensional materials. Second, transport signatures of chiral 
superconductivity have also been reported in rhombohedral 
graphene~\cite{Han2025}. Together, these experimental probes 
provide a realistic roadmap for identifying the chiral 
topological phases in the twisted graphene/NbSe$_2$ 
heterostructure and their dependence on the twist angle.

\section{Conclusions}\label{CONC}
In this work, we studied proximity-induced superconductivity in a twisted 
graphene/NbSe$_2$ van der Waals heterostructure using the Bogoliubov-de 
Gennes formalism with DFT-derived parameters of the proximitized graphene 
layer. The twist angle of $23.4^\circ$ between the graphene and NbSe$_2$ layer reduces the common symmetry of the 
heterostructure to ${\bf C}_3$, which we used to construct 
symmetry-allowed superconducting gap functions and classify them in terms of the  irreducible representations of the $\mathbf{C}_3$ group, in both the on-site and nearest-neighbor approximations.
Using these gap functions, we studied the topological phase diagram of the proximitized graphene layer as a function of the singlet-triplet and inter-triplet (singlet) mixing angles. Gap functions transforming under IR $A_0$, which preserve TRS, give $\mathbb{Z}_2 = 0$ throughout the explored parameter space, indicating the absence of a topological phase in the analyzed subspace. On the other hand, gap functions transforming according to IRs $A_{\pm 1}$, which spontaneously break TRS, give rise to a rich topological phase diagram characterized by nonzero Chern numbers. 

The physical origin of the TRS-breaking gap functions can be connected to the symmetry of the superconducting order 
parameter in monolayer NbSe$_2$ itself. In the absence of a twist, monolayer NbSe$_2$ has the ${\bf D}_{3{\rm h}}$ symmetry, with its two-dimensional IRs possessing a potentially chiral superconducting phase. The nonzero twist angle reduces the symmetry from ${\bf D}_{3{\rm h}}$ to ${\bf C}_3$, splitting the two-dimensional representation 
into the two chiral IRs, and allowing for one of them to be selected as the ground 
state. If the symmetry reduction stabilizes a TRS-breaking order 
parameter in NbSe$_2$, the proximity-induced superconductivity in 
graphene necessarily falls into the $A_1$ or $A_{-1}$ class, directly 
triggering the rich topological phase diagram analyzed in this work. The 
twist angle thus plays a dual role: it controls the strength and 
character (Rashba angle $\phi_R$ vanishes at zero twist) of the proximity-induced Rashba SOC in graphene, and simultaneously acts as a chirality selector that can potentially stabilize a TRS-breaking pairing channel. 
Our results demonstrate that the twisted graphene/NbSe$_2$ 
heterostructure is a promising and experimentally tunable platform for 
realizing and probing chiral topological superconductivity, with the twist angle providing direct experimental 
control over both the proximity-induced spin-orbit coupling and superconductivity.

\begin{acknowledgments}
Research results was obtained using the computational resources procured in the national project National competence centre for high performance computing (project code: 311070AKF2) funded by European Regional Development Fund, EU Structural Funds Informatization of society, Operational Program Integrated Infrastructure.
A.H. acknowledges the financial support provided by the Ministry of Education, Research, Development and Youth of the Slovak Republic, provided under Grant numbers APVV-21-0272 and VEGA 2/0133/25.
M.M. acknowledges the financial support by the EU NextGenerationEU through the Recovery and Resilience Plan for Slovakia under the Project No. 09I02-03-V01-00012, by the APVV grant APVV-23-0430, and VEGA grants 2/0081/26 and 2/0133/25.

\end{acknowledgments}

\appendix
\begin{figure*}[t]
\centering
\includegraphics[width=0.82\textwidth]{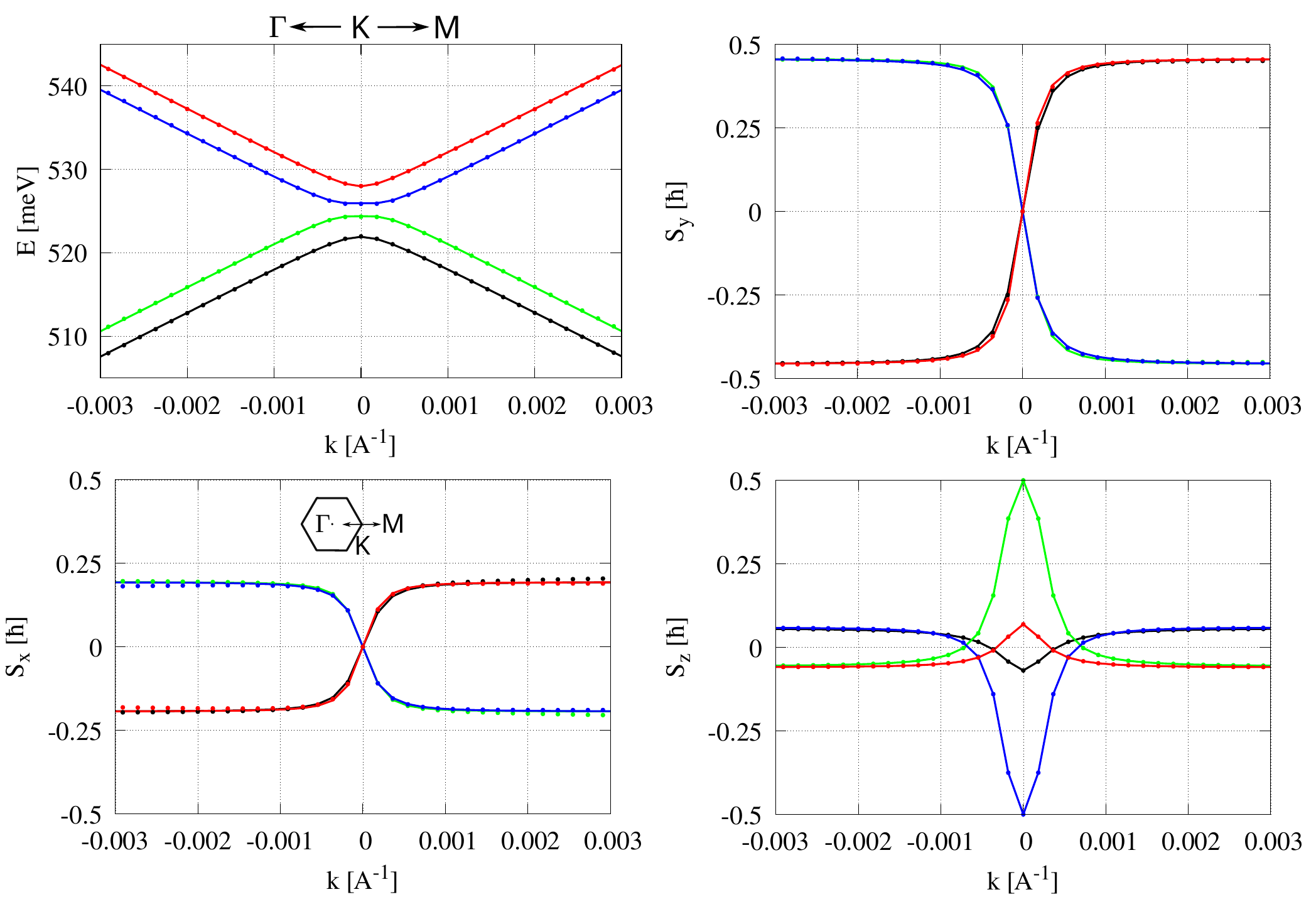}
\caption{Comparison of the electronic band structure of graphene on NbSe$_2$ and $S_x$ , $S_y$, and $S_z$ spin expectation values close to the Dirac $K$ point, with the tight-binding model Hamiltonian of graphene~\eqref{Hgr}, parametrized using the values given in Table~\ref{tab:CDW_ex}.
Whereas the solid lines are represent tight-binding model fit, the DFT data is given by circles. The k-path is along the $x$
direction ($k_y = 0$), centered around the K point.}\label{FIT}
\end{figure*} 
\section{Ab-initio calculation and effective graphene model details}\label{AppA}
Here we provide numerical details of the calculated twisted graphene/NbSe$_2$ heterostructure in its normal phase and the
fitting procedure used to obtain the effective parameters of graphene, using the model Hamiltonian~\eqref{Hgr} described into more details in the main text (Section~\ref{SectionII}).

The lattice parameter of graphene is taken as $a_0=2.46~{\rm\AA}$, while for NbSe$_2$ the latice parameter is equal to 
$a_{\rm NbSe_2}=3.48~{\rm\AA}$~\cite{DWN+11}. The supercells were constructed by straining graphene, described by the relative strain $\xi=-2.638\%$, using which the strained lattice parameter $a_0^{\rm str}$ can be described as $a_0^{\rm str}=(1+\xi[\%]/100)a_0$.  
Using the strained lattice vectors of graphene ${\bm a}_1=a_0^{\rm str}{\bm e}_x$ and ${\bm a}_2=a_0^{\rm str}(\cos{(2\pi/3)}{\bm e}_x+\sin{(2\pi/3)}{\bm e}_y)$ we can define lattice vectors of the heterostructure as $5{\bm a}_1+2{\bm a}_2$ and $-2{\bm a}_1+3{\bm a}_2$. The total number of atoms in the heterostructure is 65, consisting of 38 graphene atoms and 27 NbSe$_2$ atoms. 

We perform the electronic structure calculation of the graphene/NbSe$_2$ heterostructures using Density Functional Theory as implemented in the plane wave code Q{\sc{uantum}} ESPRESSO~\cite{QE1,QE2}. The relaxation of the studied heterostructures was performed using the Perdew-Burke-Ernzerhof functional~\cite{PBE} with projector-augmented wave~\cite{PAW} pseudopotentials.
The kinetic energy cut-offs for the wave function and charge density were chosen to be 55~Ry and 326~Ry, respectively. Additionally, Methfessel–Paxton energy level smearing~\cite{MP89} of 1~mRy was used, and $9\times 9$ $k$-points mesh for the irreducible part of the Brillouin zone sampling were used. 
The van der Waals interaction was modeled using the semiempirical Grimme-D2 correction~\cite{G06,BCF+08}, and a vacuum of 20~\AA~in the $z$-direction to detach the periodic images of the heterostructure was used. The positions of atoms were relaxed using the quasi-Newton scheme using scalar-relativistic pseudopotentials, keeping the force and energy convergence thresholds for ionic minimization to $1\times10^{-4}$~Ry/bohr and $10^{-7}$~Ry, respectively. 

For the self-consistent calculation, including the spin-orbit coupling, we used fully-relativistic pseudopotentials. Also, we have kept
the same $k$-mesh, but increased the energy convergence thresholds to $10^{-8}$~Ry. Additionally, dipole correction~\cite{B99} was applied to properly determine the Dirac point energy offset due to dipole electric field effects between graphene and NbSe$_2$.

After performing the DFT calculation of the band structure of the graphene/NbSe$_2$ heterostructure, the second step is to fit the energy bands and spin expectation values of graphene in the vicinity of the Dirac  $K=4\pi/3a_0^{\rm str}(1,0)$ point of graphene. Using the effective graphene Hamiltonian model~\eqref{Hgr}, we were able to fit the DFT data to the graphene model and extract the effective parameters. They are given in Tab.~\ref{tab:CDW_ex} of the main text, while the comparison between the fitted model and the DFT data is given in FIG.~\ref{FIT}.
We note that the $k$-path is along the $x$ direction ($k_y=0$), with the maximal distance from the $K$-point being 0.003\,\AA$^{-1}$, corresponding roughly to an  energy window [-20,20]\, meV relative to the gap center.   

\section{Determination of topological invariants via the Wilson loop calculation}
\label{app:wilson_loop}
 
The topological character of the superconducting state of proximitized graphene is determined using the Wilson loop
calculations~\cite{Yu2011,Fu2006}. Following the evolution of WFC flow across the BZ, computed from the occupied BdG eigenstates, one can calculate either a $\mathbb{Z}_2$ invariant (when TRS is preserved) or a
Chern number $C$ (when TRS is broken).

\subsection{Overlap matrix and Wannier center construction}
We analyze the BdG Hamiltonian $\mathcal{H}_{\mathrm{BdG}}(\mathbf{k})$~\eqref{EQ:BDG} 
with Bloch eigenstates $|\Psi_{n\mathbf{k}}\rangle$, where $n = 1, \ldots, 4$ 
labels the occupied BdG bands. For a fixed value of the crystal momentum 
$\mathbf{k}_1$ along $\mathbf{b}_1 = 4\pi/3\,(3/2,\sqrt{3}/2,0)$, a closed 
string of $N_2 + 1$ discretised points is constructed along the second 
reciprocal direction $\mathbf{b}_2 = 4\pi/\sqrt{3}\,(0,1,0)$,
\begin{equation}
    k_2^0,\; k_2^1,\; \ldots,\; k_2^{N_2-1},\; k_2^{N_2} \equiv k_2^0.
\end{equation}
The overlap matrix 
between two neighboring points $k_2^i$ and $k_2^{i+1}$ on this string is 
defined by its matrix elements 
\begin{equation}
    \left[M^{i,i+1}(k_1)\right]_{nm}
    = \langle\Psi_{n,\,(k_1,\,k_2^i)}|\Psi_{m,\,(k_1,\,k_2^{i+1})}\rangle,
    \label{eq:elem_overlap}
\end{equation}
$(n, m = 1, \ldots, 4)$, with $M^{i,i+1}(k_1)$  being a $4 \times 4$ matrix of inner products between occupied eigenstates 
at neighbouring $k$-points along the ${\bf b}_2$-path. At finite mesh spacing, the overlap matrices 
$M^{i,i+1}$ are not exactly unitary due to discretisation errors. Each 
overlap matrix is therefore replaced by its nearest unitary matrix in the 
Frobenius norm~\cite{Higham2008},
\begin{equation}
    \widetilde{M}^{i,i+1}(k_1)
    = \underset{U \in \mathcal{U}(4)}{\arg\min}
    \left\| M^{i,i+1}(k_1) - U \right\|_{F},
    \label{eq:nearest_unitary}
\end{equation}
which is obtained as the unitary polar factor via the singular value 
decomposition $M^{i,i+1} = X^{i,i+1}\,\Sigma^{i,i+1}\,(Y^{i,i+1})^\dagger$,
giving $\widetilde{M}^{i,i+1} = X^{i,i+1}(Y^{i,i+1})^\dagger$. This 
unitarization is a finite-mesh stabilization that does not affect the 
continuum limit of the Wilson loop. The total Wilson loop matrix at $k_1$ 
is then the ordered product of all unitarized overlap matrices around the 
closed string,
\begin{equation}
    \widetilde{M}(k_1) = \widetilde{M}^{0,1}(k_1)\,\widetilde{M}^{1,2}(k_1)
    \cdots \widetilde{M}^{N_2-1,0}(k_1),
    \label{eq:M}
\end{equation}
where the final factor closes the loop back to the starting point, due to the BZ periodicity. Using the eigenvalues
$\lambda_i(k_1) = \mathrm{eig}[ \widetilde{M}(k_1)]$, $i=1,...,4$, the WFC at $k_1$ is equal to~\cite{asboth2016short}
\begin{equation}
    \theta_i(k_1) = \frac{\arg \lambda_i(k_1)}{2\pi}
    \in \left(-\tfrac{1}{2},\, \tfrac{1}{2}\right].
    \label{eq:WCC}
\end{equation}
Repeating this procedure for each value of $k_1$ swept across the 
$\mathbf{b}_1$-path, we analyze the flow of the WFCs and read the 
topological invariant.

\begin{figure}[t]
\centering
\includegraphics[width=0.495\textwidth]{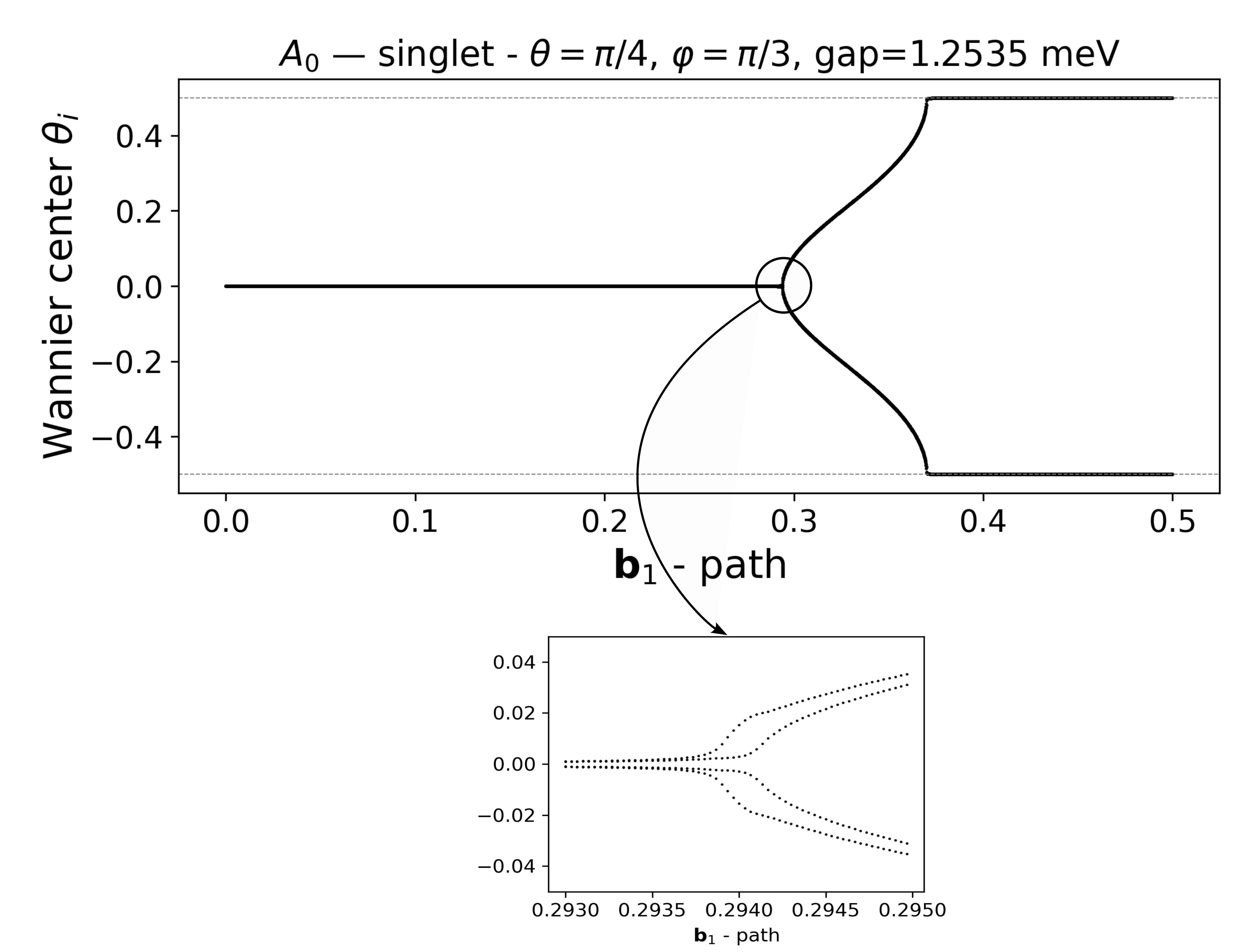}
\caption{WFC flow, computed via the Wilson loop method, for 
the singlet gap function $\Delta_{A_0}^{\rm s}(\pi/4,\pi/3)$. The main panel is obtained using the $3000\times6000$ $\mathbf{k}$-point mesh along $\mathbf{b}_1\times\mathbf{b}_2$, and suggest 
at first sight an odd number of crossings with an arbitrary horizontal line, indicating $\mathbb{Z}_2=1$. The 
insets, computed on a refined $60\times30000$ mesh, shows that the single line represent in reality two closely spaced but non-degenerate WFC lines, giving $\mathbb{Z}_2=0$. This illustrates the importance of sufficient $\mathbf{k}$-point sampling for a reliable determination of the $\mathbb{Z}_2$ invariant when the spin splitting of the electron bands is small.}\label{WLsingletAO}
\end{figure} 
\subsection{Time-reversal symmetric case: \texorpdfstring{$\mathbb{Z}_2$}{Z2} invariant}\label{Z2app}
 
When TRS is present, the WFC spectrum over the second half of the BZ is 
the mirror image of the first half. Therefore,  it is enough to compute 
the Wilson loop over half the BZ to determine the $\mathbb{Z}_2$ 
invariant~\cite{Yu2011}. The $\mathbb{Z}_2$ invariant is extracted by 
counting the number of crossings of the WFC flow with an arbitrary 
horizontal reference line $\theta_{\mathrm{ref}} \in (-\tfrac{1}{2}, 
\tfrac{1}{2})$ over the half of the BZ. The odd number of crossings, $\mathbb{Z}_2 = 1$, correspond to a topological phase, whereas the even number of crossing signals a trivial phase with $\mathbb{Z}_2=0$.

Since the gap functions transforming according to IR $A_0$ preserve TRS, one has to use the 
$\mathbb{Z}_2$ topological invariant.
The topological properties of the gap functions transforming according to the IR $A_0$ are calculated in two steps. In the first step, we consider the purely singlet gap function
$\Delta_{A_0}^{\rm s}(\theta,\phi)=\Delta_0\Big[\sin{\theta}\,
\Delta_{A_0, (s)}^{\rm os}({\phi})+\cos{\theta}\,
\Delta_{A_0, (s)}^{\rm nn}({\bf k})\Big]$, 
where $\theta$ controls the mixing between the on-site and nearest-neighbor 
singlet components, and $\phi$ parametrizes the sublattice imbalance between 
the on-site contributions on sublattices $A$ and $B$, see Eq.~\eqref{onsite}. 
Since the gap function satisfies 
$\Delta_{A_0}^{\rm s}(\theta,\phi) = -\Delta_{A_0}^{\rm s}(\theta+\pi,\phi) 
= -\Delta_{A_0}^{\rm s}(\pi-\theta,\phi+\pi)$,
and an overall sign change leaves all physical observables invariant, we 
restrict without loss of generality to $\theta\in(0,\pi/2)$, 
$\phi\in(0,2\pi)$. By evaluating $\mathbb{Z}_2$ invariant on a $3000\times6000$ 
$\mathbf{k}$-point mesh along $\mathbf{b}_1$ and $\mathbf{b}_2$ for each 
point of the $48\times192$ $(\theta,\phi)$ grid, we find $\mathbb{Z}_2=0$ 
throughout the entire parameter space. The subtlety of this determination is 
illustrated in FIG.~\ref{WLsingletAO}. In this example, the WFC flow appears at 
first to exhibit an odd number of crossings with the reference line 
$\theta_{\rm ref}$, suggesting $\mathbb{Z}_2=1$. However, a more careful 
analysis with an increased $\mathbf{k}$-point sampling along $\mathbf{b}_2$ 
reveals two closely 
spaced WFC lines, giving an odd number of crossings and 
$\mathbb{Z}_2=0$. This near-degeneracy is due to the small value of 
the Rashba SOC parameter $\lambda_{\rm R}$ in the graphene/NbSe$_2$ 
heterostructure. To verify this, we  increased 
$\lambda_{\rm R}$ and studied the evolution of the WFC spectrum. Our investigation shows that as 
$\lambda_{\rm R}$ grows, the two nearly degenerate lines become clearly 
separated without closing the gap,  confirming $\mathbb{Z}_2=0$.

After that, we investigated the effect of singlet-triplet mixing by analyzing the gap function
$\Delta_{A_0}(\theta,\phi_t)$ equal to $\Delta_{A_0}(\theta,\phi_t)=\Delta_0\Big[\cos{\theta}
\Delta_{A_0, (s)}^{\rm os}(\pi/4)+\sin{\theta}\cos{\phi_t}
\Delta_{A_0, (z)}^{\rm nn}({\bf k})+\sin{\theta}\sin{\phi_t}\,
\Delta_{A_0, (xy)}^{\rm nn}({\bf k})\Big]$, 
which describes the mixing of one on-site singlet component with equal 
sublattice contributions $\Delta_{\rm A}=\Delta_{\rm B}$ with the two triplet components 
$\Delta_{A_0,(z)}^{\rm nn}(\mathbf{k})$ and 
$\Delta_{A_0,(xy)}^{\rm nn}(\mathbf{k})$. Evaluating $\mathbb{Z}_2$ on the 
same $3000\times6000$ $\mathbf{k}$-point mesh along $\mathbf{b}_1$ and 
$\mathbf{b}_2$ for each point of the $48\times192$ $(\theta,\phi_t)$ grid, 
we again find $\mathbb{Z}_2=0$ throughout the entire parameter space. Therefore, within the two two-dimensional subspaces of the 
full parameter space explored here, we find no evidence of topological phases. We note that the full parameter space of 
$\Delta_{A_0}(\mathbf{k})$ has not been explored, and a topologically nontrivial regime may exist 
beyond the two two-dimensional subspaces analyzed here.

\begin{figure}[t]
\centering
\includegraphics[width=0.49\textwidth]{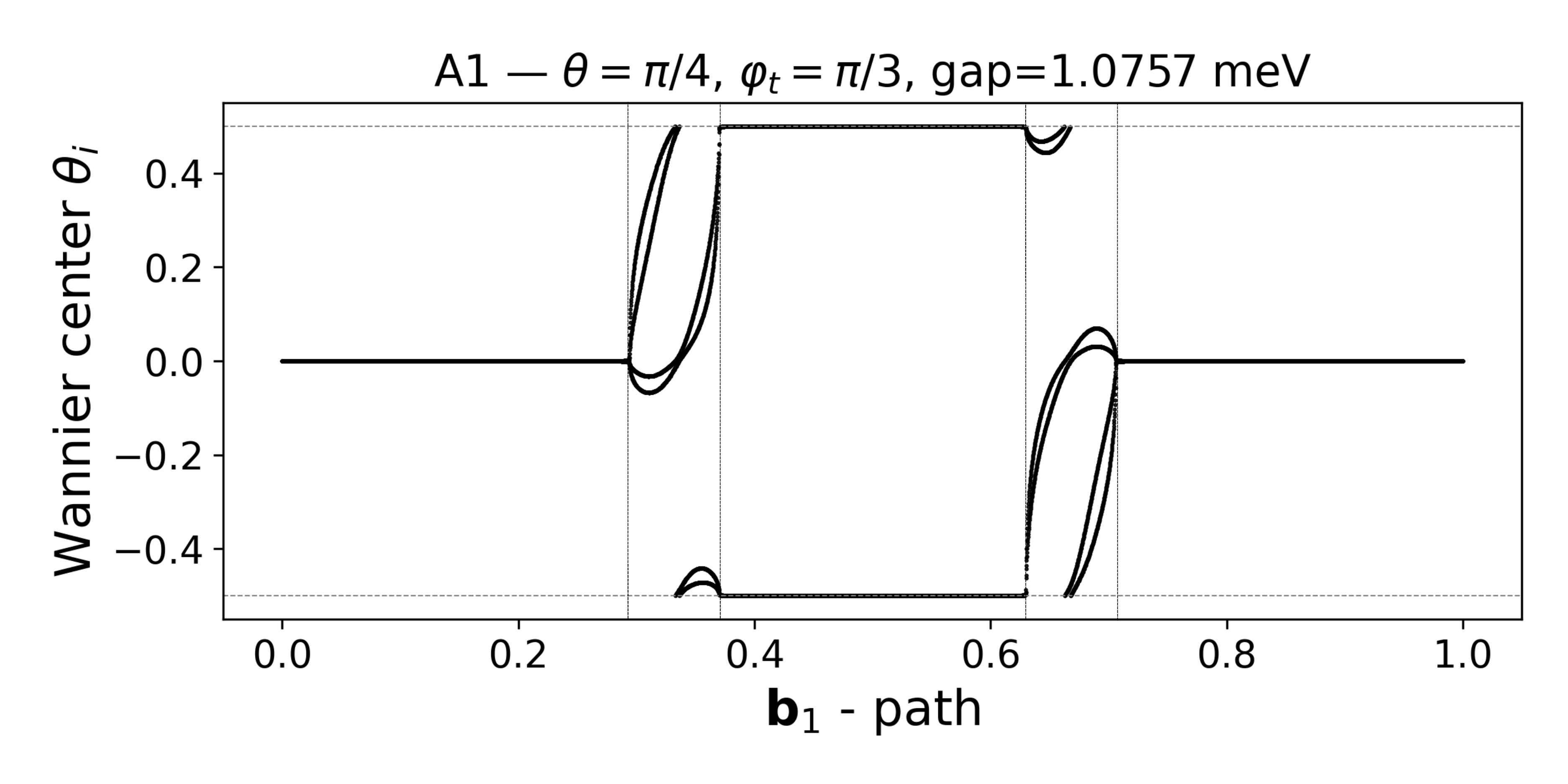}
\caption{WFC flow computed via the Wilson loop method for the gap function
$\Delta_{A_1}(\theta, \phi_t) = \Delta_{A_1}(\pi/4, \pi/3)$, obtained on a
$6000 \times 6000$ $\mathbf{k}$-point mesh, corresponding to a Chern number $C = 4$. The dashed regions, centered on $\tfrac{1}{3}\mathbf{b}_1$ and
$\tfrac{2}{3}\mathbf{b}_1$, represent parts along the $\mathbf{b}_1$-path where the
nontrivial WFC evolution appears and where a dense $\mathbf{k}$-point
sampling is needed to precisely calculate the BdG gap. }\label{WLtriplet}
\end{figure} 
\subsection{Chern number}\label{ChernNUMBER}
When TRS is broken, the relevant topological invariant is the Chern 
number $C$. The Chern number $C$ is 
extracted by counting the crossings of the WFC flow with an arbitrary 
horizontal reference line $\theta_{\rm ref} \in (-\tfrac{1}{2}, \tfrac{1}{2})$ in the full BZ. However, the crossings are slope dependent. Precisely, the WFC flow crossing $\theta_{\rm ref}$ with a positive slope counts as $+1$, and with a 
negative slope as $-1$. After calculating the intersection of all WFC branches $\theta_i$~\eqref{eq:WCC} with $\theta_{\rm ref}$, the Chern number is determined. 
Whereas $C\neq 0$ signals topologically nontrivial phase, $C = 0$ corresponds to a trivial phase.

The superconducting gap functions transforming according to IRs $A_1$ and $A_{-1}$ in~\eqref{Apm1} are related by TRS, in the sense that
applying the time-reversal operator maps the gap function
$\Delta_{A_1}(\mathbf{k})$ into $\Delta_{A_{-1}}(\mathbf{k})$. This relation directly implies $C_{A_1} =
-C_{A_{-1}}$ for the Chern numbers. Therefore, it is sufficient to analyze a single IR, which we take to be $A_1$. The corresponding gap function, which mixes all three nearest-neighbor pairing channels allowed within this IR~\eqref{Apm1}, is $\Delta_{A_{1}}({\bf k})=\Delta_0 \Big[\cos{\theta}\Delta_{A_{1}, (s)}^{\rm nn}({\bf k})+\sin{\theta}\cos{\phi_t}\Delta_{A_{ 1}, (z)}^{\rm nn}({\bf k})+\sin{\theta}\sin{\phi_t}\Delta_{A_{1 }, (xy)}^{\rm nn}({\bf k})\Big]$, where $\theta$ controls the level of singlet-triplet mixing and $\phi_t$ 
parametrizes the mixing between the two triplet components 
$\Delta_{A_1,(z)}^{\rm nn}(\mathbf{k})$ and 
$\Delta_{A_1,(xy)}^{\rm nn}(\mathbf{k})$.

Due to the numerical complexity of the Chern number calculation, the full
phase diagram in FIG.~\ref{ChernFIGURE} was obtained using a two-step procedure. This procedure is chosen to balance between the computational cost and the need to accurately determine both the BdG gap and the Chern number. In the first step, the parameter space $\theta \in (0, \pi/2)$,
$\phi_t \in (0, 2\pi)$ was sampled on a coarse $48 \times 192$
$(\theta, \phi_t)$ grid, with the topological invariants and gap values
evaluated on a uniform $6000 \times 6000$ $\mathbf{k}$-point mesh. The results reveal a rich topological
phase diagram. In particular, for $\theta \in (0, \pi/3)$ the Chern number
is equal to $C = 4$ independently of $\phi_t$, with a minimum gap of
$0.136$~meV, confirming that this region is fully gapped and topologically
nontrivial.

In the second step, the region $\theta \in (\pi/3, \pi/2)$,
$\phi_t \in (0, 2\pi)$, where gap closings and topological phase
transitions appear, was sampled on a refined $128 \times 768$
$(\theta, \phi_t)$ grid. To accurately determine the gap near $K$ and $K'$,
the gap values were computed on a uniform $18000 \times 18000$
$\mathbf{k}$-point mesh. For the
WFC flow, the full $18000$-point mesh along $\mathbf{b}_2$ was
kept, whereas the $\mathbf{b}_1$ path is divided into two strips of
$1500$ points each, centered around $\tfrac{1}{3}\mathbf{b}_1$ and
$\tfrac{2}{3}\mathbf{b}_1$ where the $K$ and $K'$ points are located and nontrivial WFC evolution is observed (see dashed regions in
FIG.~\ref{WLtriplet}). The equivalence of the gap values obtained from the
full $18000 \times 18000$ mesh and  two $1500 \times 18000$
strips confirms that both approaches sample the $K$ and $K'$ regions with
equal resolution. We note that configurations with a gap smaller than $0.04$~meV were
classified as gapless states. Also, we note that the extended phase diagram $\theta \in
(0, 2\pi)$, $\phi_t \in (0, 2\pi)$ can be reconstructed from the calculated one using the symmetry
relations $C(\pi - \theta, \phi_t + \pi) = C(\theta, \phi_t)$ and $C(\theta
+ \pi, \phi_t) = C(\theta, \phi_t)$.

\bibliography{biblio}

@article{AZ97,
  title = {Nonstandard symmetry classes in mesoscopic normal-superconducting hybrid structures},
  author = {Altland, Alexander and Zirnbauer, Martin R. },
  journal = {Phys. Rev. B},
  volume = {55},
  issue = {},
  pages = {1142},
  numpages = {},
  year = {1997},
  month = {},
  publisher = {},
  doi = {https://doi.org/10.1103/PhysRevB.55.1142},
  url ={https://journals.aps.org/prb/abstract/10.1103/PhysRevB.55.1142}}

@article{David2019,
  title = {Induced Spin-Orbit Coupling in Twisted Graphene--Transition Metal Dichalcogenide Heterobilayers: {{Twistronics}} Meets Spintronics},
  shorttitle = {Induced Spin-Orbit Coupling in Twisted Graphene--Transition Metal Dichalcogenide Heterobilayers},
  author = {David, Alessandro and Rakyta, P{\'e}ter and Korm{\'a}nyos, Andor and Burkard, Guido},
  year = {2019},
  month = aug,
  journal = {Phys. Rev. B},
  volume = {100},
  number = {8},
  pages = {085412},
  publisher = {{American Physical Society}},
  doi = {10.1103/PhysRevB.100.085412},
  url = {https://journals.aps.org/prb/abstract/10.1103/PhysRevB.100.085412},
}

@article{HFK+20,
  title = {Chiral Majorana fermions in graphene from proximity-induced superconductivity},
  author = {Högl, Petra and Frank, Tobias and Kochan, Denis and Gmitra, Martin and Fabian, Jaroslav},
  journal = {Phys. Rev. B},
  volume = {101},
  issue = {},
  pages = {245441},
  numpages = {},
  year = {2020},
  month = {},
  publisher = {},
  doi = {https://doi.org/10.1103/PhysRevB.101.245441},
  url ={https://journals.aps.org/prb/abstract/10.1103/PhysRevB.101.245441}}

@article{NGF24,
  title = {Chiral Majorana fermions in graphene from proximity-induced superconductivity},
  author = {Naimer, Thomas and Gmitra, Martin and Fabian, Jaroslav},
  journal = {Phys. Rev. B},
  volume = {109},
  issue = {},
  pages = {205109},
  numpages = {},
  year = {2024},
  month = {},
  publisher = {},
  doi = {https://doi.org/10.1103/PhysRevB.109.205109},
  url ={https://journals.aps.org/prb/abstract/10.1103/PhysRevB.109.205109}}

@book{asboth2016short,
	series = {Lecture {Notes} in {Physics}},
	title = {A {Short} {Course} on {Topological} {Insulators}},
	isbn = {978-3-319-25605-4},
	shorttitle = {A {Short} {Course} on {Topological} {Insulators}},
	volume = {909},
	DOI={10.1007/978-3-319-25607-8},
	urldate = {2019-03-01},
	publisher = {Springer International Publishing},
	author = {Asbóth, János K. and Oroszlány, László and Pályi, András},
	year = {2016},
	url={https://link.springer.com/book/10.1007/978-3-319-25607-8},
}

@article{maciejko2010topological,
  title={Topological quantization in units of the fine structure constant},
  author={Maciejko, Joseph and Qi, Xiao-Liang and Drew, H Dennis and Zhang, Shou-Cheng},
  journal={Phys. Rev. Lett.},
  volume={105},
  number={16},
  pages={166803},
  year={2010},
  publisher={APS},
  doi={https://doi.org/10.1103/PhysRevLett.105.166803},
  url={https://journals.aps.org/prl/abstract/10.1103/PhysRevLett.105.166803}
}

@article{wieder2021topological,
  title={Topological materials discovery from crystal symmetry},
  author={Wieder, Benjamin J and Bradlyn, Barry and Cano, Jennifer and Wang, Zhijun and Vergniory, Maia G and Elcoro, Luis and Soluyanov, Alexey A and Felser, Claudia and Neupert, Titus and Regnault, Nicolas and Andrei Bernevig, B.},
  journal={Nat. Rev. Mater.},
  volume={7},
  pages={196–216},
  year={2021},
  publisher={Nature Publishing Group},
  doi={https://doi.org/10.1038/s41578-021-00380-2},
  url={https://www.nature.com/articles/s41578-021-00380-2}
}

@article{HK10,
  title={Colloquium: topological insulators},
  author={Hasan, M. Z.  and Kane, C. L. },
  journal={Rev. Mod. Phys.},
  volume={82},
  number={},
  pages={3045},
  year={2010},
  publisher={APS},
  doi = {https://doi.org/10.1103/RevModPhys.82.3045},
  url ={https://journals.aps.org/rmp/abstract/10.1103/RevModPhys.82.3045}
}

@article{Yu2011,
  title = {Equivalent expression of ${\mathbb{Z}}_{2}$ topological invariant for band insulators using the non-Abelian Berry connection},
  author = {Yu, Rui and Qi, Xiao Liang and Bernevig, Andrei and Fang, Zhong and Dai, Xi},
  journal = {Phys. Rev. B},
  volume = {84},
  issue = {7},
  pages = {075119},
  numpages = {12},
  year = {2011},
  month = {Aug},
  publisher = {American Physical Society},
  doi = {10.1103/PhysRevB.84.075119},
  url = {https://link.aps.org/doi/10.1103/PhysRevB.84.075119}
}

@article{Fu2006,
  title = {Time reversal polarization and a ${Z}_{2}$ adiabatic spin pump},
  author = {Fu, Liang and Kane, C. L.},
  journal = {Phys. Rev. B},
  volume = {74},
  issue = {19},
  pages = {195312},
  numpages = {13},
  year = {2006},
  month = {Nov},
  publisher = {American Physical Society},
  doi = {10.1103/PhysRevB.74.195312},
  url = {https://link.aps.org/doi/10.1103/PhysRevB.74.195312}
}

@article{QE1,
  title = {{{QUANTUM ESPRESSO}}: A Modular and Open-Source Software Project for Quantum Simulations of Materials},
  shorttitle = {{{QUANTUM ESPRESSO}}},
  author = {Giannozzi, Paolo and Baroni, Stefano and Bonini, Nicola and Calandra, Matteo and Car, Roberto and Cavazzoni, Carlo and Ceresoli, Davide and Chiarotti, Guido L. and Cococcioni, Matteo and Dabo, Ismaila and Corso, Andrea Dal and de Gironcoli, Stefano and Fabris, Stefano and Fratesi, Guido and Gebauer, Ralph and Gerstmann, Uwe and Gougoussis, Christos and Kokalj, Anton and Lazzeri, Michele and {Martin-Samos}, Layla and Marzari, Nicola and Mauri, Francesco and Mazzarello, Riccardo and Paolini, Stefano and Pasquarello, Alfredo and Paulatto, Lorenzo and Sbraccia, Carlo and Scandolo, Sandro and Sclauzero, Gabriele and Seitsonen, Ari P. and Smogunov, Alexander and Umari, Paolo and Wentzcovitch, Renata M.},
  year = {2009},
  month = sep,
  journal = {Journal of Physics: Condensed Matter},
  volume = {21},
  number = {39},
  pages = {395502},
  doi = {10.1088/0953-8984/21/39/395502},
  langid = {english},
}

@article{QE2,
  title = {Advanced Capabilities for Materials Modelling with {{Quantum ESPRESSO}}},
  author = {Giannozzi, P. and Andreussi, O. and Brumme, T. and Bunau, O. and Nardelli, M. Buongiorno and Calandra, M. and Car, R. and Cavazzoni, C. and Ceresoli, D. and Cococcioni, M. and Colonna, N. and Carnimeo, I. and Corso, A. Dal and de Gironcoli, S. and Delugas, P. and DiStasio, R. A. and Ferretti, A. and Floris, A. and Fratesi, G. and Fugallo, G. and Gebauer, R. and Gerstmann, U. and Giustino, F. and Gorni, T. and Jia, J. and Kawamura, M. and Ko, H.-Y. and Kokalj, A. and K{\"u}{\c c}{\"u}kbenli, E. and Lazzeri, M. and Marsili, M. and Marzari, N. and Mauri, F. and Nguyen, N. L. and Nguyen, H.-V. and {Otero-de-la-Roza}, A. and Paulatto, L. and Ponc{\'e}, S. and Rocca, D. and Sabatini, R. and Santra, B. and Schlipf, M. and Seitsonen, A. P. and Smogunov, A. and Timrov, I. and Thonhauser, T. and Umari, P. and Vast, N. and Wu, X. and Baroni, S.},
  year = {2017},
  month = oct,
  journal = {Journal of Physics: Condensed Matter},
  volume = {29},
  number = {46},
  pages = {465901},
  publisher = {{IOP Publishing}},
  doi = {10.1088/1361-648X/aa8f79},
  langid = {english},
}

@article{G06,
  title = {Semiempirical GGA-type density functional constructed with a long-range dispersion correction},
  author = {Grimme, S.},
  year = {2006},
  month = aug,
  journal = {J. Comput. Chem.},
  volume = {27},
  number = {},
  pages = {1787-1799},
  publisher = {},
  doi = {https://doi.org/10.1002/jcc.20495},
  url = {https://onlinelibrary.wiley.com/doi/full/10.1002/jcc.20495},
}

@article{BCF+08,
  title = {Role and effective treatment of dispersive forces in materials: Polyethylene and graphite crystals as test cases},
  author = {Barone, V. and Casarin, M. and Forrer, D. and Pavone, M. and  Sambi, M. and Vittadini, A.},
  year = {2008},
  month = aug,
  journal = {J. Comput. Chem.},
  volume = {30},
  number = {},
  pages = {934-939},
  publisher = {},
  doi = {https://doi.org/10.1002/jcc.21112},
  url = {https://onlinelibrary.wiley.com/doi/full/10.1002/jcc.21112},
}

@article{B99,
  title = {Dipole correction for surface supercell calculations},
  author = {Bengtsson, L.},
  year = {1999},
  month = aug,
  journal = {Phys. Rev. B},
  volume = {59},
  number = {},
  pages = {12301},
  publisher = {},
  doi = {https://doi.org/10.1103/PhysRevB.59.12301},
  url = {https://journals.aps.org/prb/abstract/10.1103/PhysRevB.59.12301},
}

@article{MP89,
  title = {High-precision sampling for Brillouin-zone integration in metals},
  author = {Methfessel, M.  and Paxton, A. T.},
  year = {1989},
  month = aug,
  journal = {Phys. Rev. B},
  volume = {40},
  number = {},
  pages = {3616},
  publisher = {},
  doi = {https://doi.org/10.1103/PhysRevB.40.3616},
  url = {https://journals.aps.org/prb/abstract/10.1103/PhysRevB.40.3616},
}

@article{PAW,
  title = {Projector augmented-wave method},
  author = {Bl{\" o}chl, P. E.},
  year = {1994},
  month = aug,
  journal = {Phys. Rev. B},
  volume = {50},
  number = {},
  pages = {17953},
  publisher = {},
  doi = {https://doi.org/10.1103/PhysRevLett.77.3865},
  url = {https://journals.aps.org/prb/abstract/10.1103/PhysRevB.50.17953},
}

@article{PBE,
  title = {Generalized Gradient Approximation Made Simple},
  author = {Perdew,  J. P. and Burke,  K. and Ernzerhof, M.},
  year = {1996},
  month = aug,
  journal = {Phys. Rev. Lett.},
  volume = {77},
  number = {},
  pages = {3865},
  publisher = {},
  doi = {https://doi.org/10.1103/PhysRevLett.77.3865},
  url = {https://journals.aps.org/prl/abstract/10.1103/PhysRevLett.77.3865},
}

@article{DWN+11,
  title = {First principles study of structural, vibrational and electronic properties of graphene-like MX2 (M=Mo, Nb, W, Ta; X=S, Se, Te) monolayers},
  author = {Ding, Yi and Wang, Yanli and Ni, Jun and Shi, Lin and Shi, Siqi and Tang, Weihua },
  year = {2011},
  journal = {Phys. B: Condens. Matter},
  volume = {406},
  number = {},
  pages = {2254},
  publisher = {},
  doi = {https://doi.org/10.1016/j.physb.2011.03.044},
  url = {https://www.sciencedirect.com/science/article/pii/S0921452611002651?via%3Dihub},
}

@article{ZFS04,
  title = {Spintronics: Fundamentals and applications},
  author = {{\v Z}uti{\' c}, Igor and Fabian, Jaroslav and Das Sarma, Sancar},
  year = {2004},
  month = apr,
  journal = {Review of Modern Physics},
  volume = {76},
  number = {},
  pages = {323},
  publisher = {American Physical Society},
  doi = {10.1103/RevModPhys.76.323},
}

@article{FME+07,
  title = {Semiconductor Spintronics},
  author = {Fabian, J. and Matos-Abiague,  A. and Ertler,  C. and Stano, P. and {\v Z}uti{\' c}, I.},
  year = {2007},
  month = aug,
  journal = {Acta Phys. Slovaca},
  volume = {57},
  number = {},
  pages = {565},
  publisher = {},
  doi = {https://doi.org/10.48550/arXiv.0711.1461},
}

@article{NGM+04,
  title = {Electric Field Effect in Atomically Thin Carbon Films},
  author = {Novoselov, K. S. and Geim, A. K. and Morozov, S. V. and Jiang, D. and Zhang, Y. and  Dubonos, S. V. and Grigorieva, I. V. and Firsov, A. A.},
  year = {2004},
  journal = {Science},
  volume = {306},
  number = {},
  pages = {666},
  publisher = {},
  doi = {https://doi.org/10.1126/science.1102896},
  url = {https://www.science.org/doi/10.1126/science.1102896},
}

@article{HGN+06,
  title = {Graphene Spin Valve Devices},
  author = {Hill, E. W. and Geim, A. K. and  Novoselov, K. and Schedin,  F. and Blake, P.},
  year = {2006},
  journal = {IEEE Trans. Magn.},
  volume = {42},
  number = {},
  pages = {2694-2696},
  publisher = {},
  doi = {10.1109/TMAG.2006.878852},
  url = {https://doi.org/10.1109/TMAG.2006.878852},
}

@article{TJP+07,
  title = {Electronic spin transport and spin precession in single graphene layers at room temperature},
  author = {Tombros, Nikolaos and Jozsa, Csaba and Popinciuc, Mihaita and Jonkman, Harry T. and van Wees, Bart J. },
  year = {2007},
  journal = {Nature},
  volume = {448},
  number = {},
  pages = {571–574},
  publisher = {},
  doi = {https://doi.org/10.1038/nature06037},
  url = {https://www.nature.com/articles/nature06037},
}

@article{OSN+07,
  title = {Spin injection into a graphene thin film at room temperature},
  author = {Ohishi, M. and Shiraishi, M. and Nouchi, R. and  Nozaki, T. and Shinjo, T. and Suzuki, Y. },
  year = {2007},
  journal = {Jpn. J. Appl. Phys.},
  volume = {46},
  number = {},
  pages = {L605},
  publisher = {},
  doi = {10.1143/JJAP.46.L605},
  url = {https://iopscience.iop.org/article/10.1143/JJAP.46.L605},
}

@article{NMA+25,
  title = {Superconductivity controlled by twist angle in monolayer NbSe2 on graphene},
  author = {Naritsuka, Masahiro and Machida, Tadashi and Asano, Shun and Yanase, Youichi and Hanaguri, Tetsuo},
  year = {2025},
  journal = {Nat. Phys.},
  volume = {21},
  number = {},
  pages = {746–753},
  publisher = {},
  doi = {https://doi.org/10.1038/s41567-025-02828-6},
  url = {https://www.nature.com/articles/s41567-025-02828-6},
}

@article{TKO+25,
  title = {From Local to Collective Superconductivity in Proximitized Graphene},
  author = {Trivini, Stefano and Kokkeler, Tim and Ortuzar, Jon and Cortés-del Río, Eva and Viña-Bausá, Beatriz and Mallet, Pierre and Veuillen, Jean-Yves  and   Cuevas, Juan Carlos and  Brihuega, Ivan and Bergeret, F. Sebastian and Pascual, Jose Ignacio},
  year = {2025},
  journal = {Nano Lett.},
  volume = {25},
  number = {46},
  pages = {16323–16329},
  publisher = {},
  doi = {https://doi.org/10.1021/acs.nanolett.5c03487},
  url = {https://pubs.acs.org/doi/10.1021/acs.nanolett.5c03487},
}

@article{Sanchez1995,
  title     = {Specific heat of 2H-{NbSe$_2$} in high magnetic fields},
  author    = {Sanchez, D. and Junod, A. and Muller, J. and Berger, H. and Levy, F.},
  journal   = {Physica B},
  volume    = {204},
  number    = {1-4},
  pages     = {167--175},
  year      = {1995},
  publisher = {Elsevier},
  doi       = {https://doi.org/10.1016/0921-4526(94)00259-X},
  url       = {https://www.sciencedirect.com/science/article/pii/092145269400259X}
}

@article{Xi2016,
  title     = {Ising pairing in superconducting {NbSe$_2$} atomic layers},
  author    = {Xi, X. and Wang, Z. and Zhao, W. and Park, J.-H. and Law, K. T.
               and Berger, H. and Forr\'o, L. and Shan, J. and Mak, K. F.},
  journal   = {Nat. Phys.},
  volume    = {12},
  pages     = {139},
  year      = {2016},
  publisher = {Nature Publishing Group},
  doi       = {https://doi.org/10.1038/nphys3538},
  url       = {https://www.nature.com/articles/nphys3538}
}

@article{Khestanova2018,
  title     = {Unusual suppression of the superconducting energy gap and critical
               temperature in atomically thin {NbSe$_2$}},
  author    = {Khestanova, E. and Birkbeck, J. and Zhu, M. and Cao, Y. and Yu, G.
               and Ghazaryan, D. and Yin, J. and Berger, H. and Forro, L.
               and Taniguchi, T. and Watanabe, K. and Novoselov, K. S.
               and Mishchenko, A. and Gorbachev, R. and Geim, A. K.
               and Grigorieva, I. V.},
  journal   = {Nano Lett.},
  volume    = {18},
  pages     = {2623},
  year      = {2018},
  publisher = {ACS},
  doi       = {https://doi.org/10.1021/acs.nanolett.8b00443},
  url       = {https://pubs.acs.org/doi/10.1021/acs.nanolett.8b00443}
}

@article{Dai2017,
  title     = {Proximity-effect-induced superconducting gap in topological surface
               states -- a point contact spectroscopy study of
               {NbSe$_2$/Bi$_2$Se$_3$} superconductor--topological insulator
               heterostructures},
  author    = {Dai, W. and Richardella, A. and Du, R. and Zhao, W. and Liu, X.
               and Liu, C. X. and Huang, S.-H. and Sankar, R. and Chou, F.
               and Samarth, N. and Li, Q.},
  journal   = {Sci. Rep.},
  volume    = {7},
  pages     = {7631},
  year      = {2017},
  publisher = {Nature Publishing Group},
  doi       = {https://doi.org/10.1038/s41598-017-07990-3},
  url       = {https://www.nature.com/articles/s41598-017-07990-3}
}

@article{Hanis2024,
  title     = {Distinguishing nodal and nonunitary superconductivity in quasiparticle 
               interference of an Ising superconductor with Rashba spin-orbit coupling: 
               The example of {NbSe$_2$}},
  author    = {Hani\v{s}, J. and Milivojevi\'{c}, M. and Gmitra, M.},
  journal   = {Phys. Rev. B},
  volume    = {110},
  pages     = {104502},
  year      = {2024},
  publisher = {American Physical Society},
  doi       = {10.1103/PhysRevB.110.104502},
  url       = {https://link.aps.org/doi/10.1103/PhysRevB.110.104502}
}

@article{Milivojevic2026,
  title     = {Machine learning protocol to identify pairing symmetries via 
               quasiparticle interference imaging in Ising superconductors},
  author    = {Hlo\v{z}n\'{y}, A. and Hani\v{s}, J. and Gmitra, M. and 
               Milivojevi\'{c}, M.},
  journal   = {arXiv preprint},
  year      = {2026},
  eprint    = {2602.19791},
  archivePrefix = {arXiv},
  primaryClass  = {cond-mat},
  url       = {https://arxiv.org/abs/2602.19791}
}

@article{Siegl2025,
  title     = {Friedel oscillations and chiral superconductivity in monolayer 
               {NbSe$_2$}},
  author    = {Siegl, J. and Bleibaum, A. and Wan, W. and Kurpas, M. 
               and Schliemann, J. and Ugeda, M.~M. and Marganska, M. 
               and Grifoni, M.},
  journal   = {Nat. Commun.},
  volume    = {16},
  pages     = {8228},
  year      = {2025},
  publisher = {Nature Publishing Group},
  doi       = {10.1038/s41467-025-63319-z},
  url       = {https://www.nature.com/articles/s41467-025-63319-z}
}

@article{Engstrom2025,
  title     = {Upper Critical Field and Pairing Symmetry of Ising Superconductors},
  author    = {Engstr\"{o}m, L. and Zullo, L. and Cren, T. and Mesaros, A. 
               and Simon, P.},
  journal   = {Phys. Rev. Lett.},
  volume    = {135},
  pages     = {236004},
  year      = {2025},
  publisher = {American Physical Society},
  doi       = {10.1103/bnw9-xx7v},
  url       = {https://journals.aps.org/prl/abstract/10.1103/bnw9-xx7v}
}

@article{Offidani2017,
  title     = {Optimal charge-to-spin conversion in graphene on transition  metal dichalcogenides},
  author    = {Offidani, M. and Milletar\`{i}, M. and Raimondi, R. 
               and Ferreira, A.},
  journal   = {Phys. Rev. Lett.},
  volume    = {119},
  pages     = {196801},
  year      = {2017},
  publisher = {American Physical Society},
  doi       = {10.1103/PhysRevLett.119.196801}
}

@article{Ghiasi2019,
  title     = {Charge-to-spin conversion by the Rashba-Edelstein effect in two-dimensional van der Waals heterostructures up to room temperature},
  author    = {Ghiasi, T.~S. and Kaverzin, A.~A. and Blah, P.~J. and van Wees, B.~J.},
  journal   = {Nano Lett.},
  volume    = {19},
  pages     = {5959-5966},
  year      = {2019},
  publisher = {American Chemical Society},
  doi       = {10.1021/acs.nanolett.9b01611},
  url       = {https://pubs.acs.org/doi/10.1021/acs.nanolett.9b01611}
}

@article{Sinova2015,
  title     = {Spin Hall effects},
  author    = {Sinova, J. and Valenzuela, S.~O. and Wunderlich, J. and Back, C.~H. and Jungwirth, T.},
  journal   = {Rev. Mod. Phys.},
  volume    = {87},
  pages     = {1213},
  year      = {2015},
  publisher = {American Physical Society},
  doi       = {10.1103/RevModPhys.87.1213},
  url       = {https://link.aps.org/doi/10.1103/RevModPhys.87.1213}
}

@article{GOW17,
  title     = {Bias induced up to 100\% spin-injection and detection polarizations 
               in ferromagnet/bilayer-{hBN}/graphene/{hBN} heterostructures},
  author    = {Gurram, M. and Omar, S. and van Wees, B.~J.},
  journal   = {Nat. Commun.},
  volume    = {8},
  pages     = {248},
  year      = {2017},
  publisher = {Nature Publishing Group},
  doi       = {10.1038/s41467-017-00317-w},
  url       = {https://www.nature.com/articles/s41467-017-00317-w}
}

@article{GOW18,
  title     = {Electrical spin injection, transport, and detection in 
               graphene-hexagonal boron nitride van der {W}aals heterostructures: 
               progress and perspectives},
  author    = {Gurram, M. and Omar, S. and van Wees, B.~J.},
  journal   = {2D Mater.},
  volume    = {5},
  pages     = {032004},
  year      = {2018},
  publisher = {IOP Publishing},
  doi       = {10.1088/2053-1583/aac34c},
  url       = {https://iopscience.iop.org/article/10.1088/2053-1583/aac34c}
}

@article{Popinciuc2009,
  title     = {Electronic spin transport in graphene field-effect transistors},
  author    = {Popinciuc, M. and J\'{o}zsa, C. and Zomer, P.~J. and Tombros, N. and Veligura, A. and Jonkman, H.~T. and van Wees, B.~J.},
  journal   = {Phys. Rev. B},
  volume    = {80},
  pages     = {214427},
  year      = {2009},
  publisher = {American Physical Society},
  doi       = {10.1103/PhysRevB.80.214427},
  url       = {https://link.aps.org/doi/10.1103/PhysRevB.80.214427}
}

@article{Ringer2018,
  title     = {Spin field-effect transistor action via tunable polarization of the spin injection in a {Co/MgO/graphene} contact},
  author    = {Ringer, S. and Rosenauer, M. and V\"{o}lkl, T. and Kadur, M. and Hopperdietzel, F. and Weiss, D. and Eroms, J.},
  journal   = {Appl. Phys. Lett.},
  volume    = {113},
  pages     = {132403},
  year      = {2018},
  publisher = {AIP Publishing},
  doi       = {https://doi.org/10.1063/1.5049664},
  url       = {https://pubs.aip.org/aip/apl/article-abstract/113/13/132403/35478/Spin-field-effect-transistor-action-via-tunable?redirectedFrom=fulltext}
}

@article{Avsar2014,
  title     = {Spin-orbit proximity effect in graphene},
  author    = {Avsar, A. and Tan, J.~Y. and Taychatanapat, T. and Balakrishnan, J. and Koon, G.~K.~W. and Yeo, Y. and Lahiri, J. and Carvalho, A. and Rodin, A.~S. and O'Farrell, E.~C.~T. and Eda, G. and Castro~Neto, A.~H. and \"{O}zyilmaz, B.},
  journal   = {Nat. Commun.},
  volume    = {5},
  pages     = {4875},
  year      = {2014},
  publisher = {Nature Publishing Group},
  doi       = {10.1038/ncomms5875},
  url       = {https://www.nature.com/articles/ncomms5875}
}

@article{Gmitra2015,
  title     = {Graphene on transition-metal dichalcogenides: {A} platform for proximity spin-orbit physics and optospintronics},
  author    = {Gmitra, M. and Fabian, J.},
  journal   = {Phys. Rev. B},
  volume    = {92},
  pages     = {155403},
  year      = {2015},
  publisher = {American Physical Society},
  doi       = {10.1103/PhysRevB.92.155403},
  url       = {https://link.aps.org/doi/10.1103/PhysRevB.92.155403}
}

@article{YTL+16,
  title     = {A two-dimensional spin field-effect switch},
  author    = {Yan, W. and Txoperena, O. and Llopis, R. and Dery, H. and Hueso, L.~E. and Casanova, F.},
  journal   = {Nat. Commun.},
  volume    = {7},
  pages     = {13372},
  year      = {2016},
  publisher = {Nature Publishing Group},
  doi       = {10.1038/ncomms13372},
  url       = {https://www.nature.com/articles/ncomms13372}
}

@article{CSC+22,
  title     = {Reliability of spin-to-charge conversion measurements in 
               graphene-based lateral spin valves},
  author    = {Safeer, C.~K. and Herling, F. and Choi, W.~Y. and Ontoso, N. 
               and Ingla-Ayn\'{e}s, J. and Hueso, L.~E. and Casanova, F.},
  journal   = {2D Mater.},
  volume    = {9},
  pages     = {015024},
  year      = {2022},
  publisher = {IOP Publishing},
  doi       = {10.1088/2053-1583/ac3c9b},
  url       = {https://iopscience.iop.org/article/10.1088/2053-1583/ac3c9b}
}

@article{Lee2022b,
  title     = {Charge-to-spin conversion in twisted graphene/{WSe$_2$} 
               heterostructures},
  author    = {Lee, S. and de~Sousa, D.~J.~P. and Kwon, Y.-K. and de~Juan, F. 
               and Chi, Z. and Casanova, F. and Low, T.},
  journal   = {Phys. Rev. B},
  volume    = {106},
  pages     = {165420},
  year      = {2022},
  publisher = {American Physical Society},
  doi       = {10.1103/PhysRevB.106.165420},
  url       = {https://link.aps.org/doi/10.1103/PhysRevB.106.165420}
}

@article{OSH+23,
  title     = {Unconventional charge-to-spin conversion in graphene/{MoTe$_2$} 
               van der {W}aals heterostructures},
  author    = {Ontoso, N. and Safeer, C.~K. and Herling, F. and Ingla-Ayn\'{e}s, J. 
               and Yang, H. and Chi, Z. and Martin-Garcia, B. and Robredo, I. 
               and Vergniory, M.~G. and de~Juan, F. and Calvo, M.~R. 
               and Hueso, L.~E. and Casanova, F.},
  journal   = {Phys. Rev. Applied},
  volume    = {19},
  pages     = {014053},
  year      = {2023},
  publisher = {American Physical Society},
  doi       = {10.1103/PhysRevApplied.19.014053},
  url       = {https://link.aps.org/doi/10.1103/PhysRevApplied.19.014053}
}

@article{Chi2024,
  title     = {Control of charge-spin interconversion in van der {W}aals 
               heterostructures with chiral charge density waves},
  author    = {Chi, Z. and Lee, S. and Yang, H. and Dolan, E. and Safeer, C.~K. and Ingla-Ayn\'{e}s, J. and Herling, F. and Ontoso, N. and Mart\'{i}n-Garc\'{i}a, B. and Gobbi, M. and Low, T. and Hueso, L.~E. and Casanova, F.},
  journal   = {Adv. Mater.},
  volume    = {36},
  pages     = {2310768},
  year      = {2024},
  publisher = {Wiley},
  doi       = {10.1002/adma.202310768},
  url       = {https://onlinelibrary.wiley.com/doi/10.1002/adma.202310768}
}

@article{CDY+25,
  title     = {Gate-tunable charge-spin interconversion in graphene/heavy-metal 
               heterostructures},
  author    = {Chi, Z. and Dolan, E. and Yang, H. and Mart\'{i}n-Garc\'{i}a, B. 
               and Gobbi, M. and Hueso, L.~E. and Casanova, F.},
  journal   = {Phys. Rev. Applied},
  volume    = {24},
  pages     = {064076},
  year      = {2025},
  publisher = {American Physical Society},
  doi       = {10.1103/k5tg-xm6q},
  url       = {https://link.aps.org/doi/10.1103/k5tg-xm6q}
}

@article{MMJ+26,
  title     = {Ferroelectric switching control of spin current in graphene 
               proximitized by {In$_2$Se$_3$}},
  author    = {Milivojevi\'{c}, M. and Mnich, J. and Jureczko, P. 
               and Kurpas, M. and Gmitra, M.},
  journal   = {Mater. Futures},
  volume    = {5},
  pages     = {015201},
  year      = {2026},
  publisher = {IOP Publishing},
  doi       = {10.1088/2752-5724/ae18ea},
  url       = {https://iopscience.iop.org/article/10.1088/2752-5724/ae18ea}
}

@article{ALR+24,
  title     = {Colloquium: {S}pin-orbit effects in superconducting hybrid 
               structures},
  author    = {Amundsen, M. and Linder, J. and Robinson, J.~W.~A. 
               and \v{Z}uti\'{c}, I. and Banerjee, N.},
  journal   = {Rev. Mod. Phys.},
  volume    = {96},
  pages     = {021003},
  year      = {2024},
  publisher = {American Physical Society},
  doi       = {10.1103/RevModPhys.96.021003},
  url       = {https://link.aps.org/doi/10.1103/RevModPhys.96.021003}
}

@article{MHL+20,
  title     = {Interfacial spin-orbit coupling: {A} platform for 
               superconducting spintronics},
  author    = {Mart\'{i}nez, I. and H\"{o}gl, P. and Gonz\'{a}lez-Ruano, C. 
               and Cascales, J.~P. and Tiusan, C. and Lu, Y. and Hehn, M. 
               and Matos-Abiague, A. and Fabian, J. and \v{Z}uti\'{c}, I. 
               and Aliev, F.~G.},
  journal   = {Phys. Rev. Applied},
  volume    = {13},
  pages     = {014030},
  year      = {2020},
  publisher = {American Physical Society},
  doi       = {10.1103/PhysRevApplied.13.014030},
  url       = {https://link.aps.org/doi/10.1103/PhysRevApplied.13.014030}
}

@article{ZPT+23,
  title     = {Enhanced superconductivity in spin-orbit proximitized bilayer 
               graphene},
  author    = {Zhang, Y. and Polski, R. and Thomson, A. and 
               Lantagne-Hurtubise, \'{E}. and Lewandowski, C. and Zhou, H. 
               and Watanabe, K. and Taniguchi, T. and Alicea, J. 
               and Nadj-Perge, S.},
  journal   = {Nature},
  volume    = {613},
  pages     = {268-273},
  year      = {2023},
  publisher = {Springer Nature},
  doi       = {10.1038/s41586-022-05446-x},
  url       = {https://www.nature.com/articles/s41586-022-05446-x}
}

@article{ZSM+25,
  title     = {Twist-programmable superconductivity in spin-orbit-coupled bilayer graphene},
  author    = {Zhang, Y. and Shavit, G. and Ma, H. and Han, Y. and Siu, C.~W. 
and Mukherjee, A. and Watanabe, K. and Taniguchi, T. and Hsieh, D. and Lewandowski, C. and von~Oppen, F. and Oreg, Y. and Nadj-Perge, S.},
  journal   = {Nature},
  volume    = {641},
  pages     = {625-631},
  year      = {2025},
  publisher = {Springer Nature},
  doi       = {10.1038/s41586-025-08959-3},
  url       = {https://www.nature.com/articles/s41586-025-08959-3}
}

@article{CLA+22,
  title     = {Nodal and nematic superconducting phases in {NbSe$_2$} monolayers from competing superconducting channels},
  author    = {Cho, C.-W. and Lyu, J. and An, L. and Han, T. and Lo, K.~T. and Ng, C.~Y. and Hu, J. and Gao, Y. and Li, G. and Huang, M. and Wang, N. and Schmalian, J. and Lortz, R.},
  journal   = {Phys. Rev. Lett.},
  volume    = {129},
  pages     = {087002},
  year      = {2022},
  publisher = {American Physical Society},
  doi       = {10.1103/PhysRevLett.129.087002},
  url       = {https://link.aps.org/doi/10.1103/PhysRevLett.129.087002}
}

@article{Kochan2017,
  title = {Model Spin-Orbit Coupling {{Hamiltonians}} for Graphene Systems},
  author = {Kochan, Denis and Irmer, Susanne and Fabian, Jaroslav},
  year = {2017},
  month = apr,
  journal = {Phys. Rev. B},
  volume = {95},
  number = {16},
  pages = {165415},
  publisher = {{American Physical Society}},
  doi = {10.1103/PhysRevB.95.165415},
  url ={https://journals.aps.org/prb/abstract/10.1103/PhysRevB.95.165415},
}

@article{Chiral_QPI2025,
  title     = {Microscopic fingerprint of chiral superconductivity},
  author    = {Wu, X. and Hao, X. and Chen, Z. and Cai, Y. and Wu, M. and Chen, C. and Wang, K. and Ming, F. and Johnston, S. and Zhang, R.-X. and Weitering, H.~H.},
  journal   = {Phys. Rev. X},
  volume    = {16},
  pages     = {011026},
  year      = {2026},
  doi       = {10.1103/jmmf-mpr8},
  url       = {https://journals.aps.org/prx/abstract/10.1103/jmmf-mpr8}
}

@article{Han2025,
  title     = {Signatures of chiral superconductivity in rhombohedral graphene},
  author    = {Han, T. and Lu, Z. and Hadjri, Z. and Shi, L. and Wu, Z. 
 and Xu, W. and Yao, Y. and Cotten, A.~A. and Sharifi~Sedeh, O. and Weldeyesus, H. and Yang, J. and Seo, J. and Ye, S. and Zhou, M. and Liu, H. and Shi, G. and Hua, Z. and Watanabe, K. and Taniguchi, T. and Xiong, P. and Zumb\"{u}hl, D.~M. and Fu, L. and Ju, L.},
  journal   = {Nature},
  volume    = {643},
  pages     = {654-661},
  year      = {2025},
  publisher = {Springer Nature},
  doi       = {https://doi.org/10.1038/s41586-025-09169-7},
  url       = {https://doi.org/10.1038/s41586-025-09169-7}
}

@article{Pangburn2023,
  title     = {Superconductivity in monolayer and few-layer graphene. 
               {I}. {R}eview of possible pairing symmetries and basic 
               electronic properties},
  author    = {Pangburn, E. and Haurie, L. and Cr\'{e}pieux, A. 
               and Awoga, O.~A. and Black-Schaffer, A.~M. 
               and P\'{e}pin, C. and Bena, C.},
  journal   = {Phys. Rev. B},
  volume    = {108},
  pages     = {134514},
  year      = {2023},
  publisher = {American Physical Society},
  doi       = {10.1103/PhysRevB.108.134514},
  url       = {https://link.aps.org/doi/10.1103/PhysRevB.108.134514}
}

@book{Higham2008,
  title     = {Functions of Matrices: Theory and Computation},
  author    = {Higham, N.~J.},
  year      = {2008},
  publisher = {SIAM},
  address   = {Philadelphia},
  doi       = {10.1137/1.9780898717778},
  url       = {https://epubs.siam.org/doi/book/10.1137/1.9780898717778}
}

@article{FHS05,
  title     = {Chern Numbers in Discretized {B}rillouin Zone: 
               Efficient Method of Computing {(S}pin{)} Hall Conductances},
  author    = {Fukui, T. and Hatsugai, Y. and Suzuki, H.},
  journal   = {J. Phys. Soc. Jpn.},
  volume    = {74},
  pages     = {1674-1677},
  year      = {2005},
  publisher = {Physical Society of Japan},
  doi       = {10.1143/JPSJ.74.1674},
  url       = {https://journals.jps.jp/doi/10.1143/JPSJ.74.1674}
}

@misc{Tong,
  title        = {Gauge Theory},
  author       = {Tong, D.},
  year         = {2018},
  howpublished = {Lecture notes, University of Cambridge},
  url          = {https://www.damtp.cam.ac.uk/user/tong/gaugetheory.html}
}
\end{document}